%% file: main.tex
\documentclass[notitlepage,aps,prd,twocolumn,superscriptaddress,showpacs,nofootinbib]{revtex4-2}

\usepackage{natbib}
\usepackage[english]{babel}
\usepackage[utf8]{inputenc}
\DeclareUnicodeCharacter{2009}{\,}

\usepackage{enumitem}
\usepackage{braket}
\usepackage{hyperref}
\usepackage{footnote}
\usepackage{tablefootnote}
\usepackage{amsmath,amssymb,amsfonts,mathrsfs}
\usepackage{booktabs}
\usepackage{mathtools}
\usepackage{bm}
\usepackage{dcolumn}
\usepackage{graphicx}   
\usepackage{latexsym}
\usepackage{color}
\usepackage[acronym]{glossaries}
\usepackage{physics}
\usepackage{siunitx}

\newcommand\be{\begin{equation}}
\newcommand\ba{\begin{eqnarray}}
\newcommand\ee{\end{equation}}
\newcommand\ea{\end{eqnarray}}

\newcommand{\RNum}[1]{\uppercase\expandafter{\romannumeral #1\relax}}

\newcommand{\ddd}{\mathrm{d}}

\DeclareSIUnit{\persqrthz}{\ensuremath{\text{Hz}^{-1/2}}}

\allowdisplaybreaks

\begin{document}

\title {Observing Kinematic Anisotropies of the Stochastic Background  with LISA }

\author{Lavinia Heisenberg}
\email{heisenberg@thphys.uni-heidelberg.de}
\affiliation{Institute for Theoretical Physics, University of Heidelberg, Philosophenweg 16
D-69120	Heidelberg
Germany} 

\author{Henri Inchausp\'e}
\email{h.inchauspe@thphys.uni-heidelberg.de}
\affiliation{Institute for Theoretical Physics, University of Heidelberg, Philosophenweg 16
D-69120	Heidelberg
Germany}

\author{David Maibach}
\email{d.maibach@thphys.uni-heidelberg.de}
\affiliation{Institute for Theoretical Physics, University of Heidelberg, Philosophenweg 16
D-69120	Heidelberg
Germany} 

\date{\today}

\begin{abstract}
We propose a diagnostic tool for future analyses of stochastic gravitational wave background signals of extra-galactic origin in LISA data. Next-generation gravitational wave detectors hold the capability to track unresolved gravitational waves bundled into a stochastic background. This composite background contains cosmological and astrophysical contributions, the exploration of which offers promising avenues for groundbreaking new insights into very early universe cosmology as well as late-time structure formation. In this article, we develop a full end-to-end pipeline for the extraction of extra-galactic signals, based on kinematic anisotropies arising from the galactic motion, via full-time-domain simulations of LISA's response to the gravitational wave anisotropic sky. Employing a Markov-Chain-Monte-Carlo map-making scheme, multipoles up to $\ell=2$ are recovered for scale-free spectra that support an interpretation as signals originating from cosmic strings in the case of a high signal-to-noise ratio. We demonstrate that our analysis is consistently beating cosmic variance and is robust against statistical and systematic errors. The impact of instrumental noise on the extraction of kinematic anisotropies is investigated, and we establish a detection threshold of $\Omega_{GW}\gtrsim 5\times 10^{-8}$ in the presence of instrument-induced noise. Potential avenues for improvement in our methodology are highlighted.

\end{abstract}


\maketitle


\newcommand{\myhyperref}[1]{\hyperref[#1]{\ref{#1}}}


\input{acronym}

\section{Introduction} 
\label{sec:intro}
The continuous enhancement of \gls{gw} measurements has ushered in a new era in GW astronomy. While the initial detection efforts focused on resolved waveforms, such as the first measurement by the LIGO/Virgo collaboration in 2015 \cite{ligo_scientific_collaboration_and_virgo_collaboration_binary_2016}, we are now approaching the capability of detecting unresolved \gls{gw} sources with upcoming probes \cite{baghi_uncovering_2023, hartwig_characterization_2022,  caprini_cosmological_2018}. The sheer number of unresolved sources is so vast that they collectively create a stochastic signal known as the \gls{sgwb}. This background encompasses a wide array of contributions from astrophysical as well as cosmological sources (see \cite{caprini_cosmological_2018} and related references for a comprehensive review). Rich in its phenomenology, the \gls{sgwb} can serve as compelling evidence for new physics on cosmological scales, offering a distinct avenue of exploration beyond high-energy physics at the TeV scale and probing the early universe prior to the \gls{cmb}.\\
Very recently, Pulsar Timing Arrays (PTAs) \cite{Maiorano_2021} have managed to collect initial evidence of a stochastic background \cite{agazie_nanograv_2023_II, afzal_nanograv_2023_I} flooding the universe. Despite the origins of this background remaining uncertain to this day, the results of PTA represents the initial step in identifying the astrophysical contributions to the \gls{sgwb}. With forthcoming space-based instruments like the \gls{lisa} \cite{amaro-seoane_laser_2017}, insights gained from PTA can be supported by additional data potentially enhancing the current upper limits on the \gls{sgwb} from ground-based instruments \cite{LV_2009,abbott_2021_limit, renzini2022stochastic}.
In addition, future ground-based instruments, such as the Einstein Telescope \cite{Branchesi_2023, Maggiore_2020} and Cosmic Explorer \cite{evans2023cosmic, reitze2019cosmic}, will play a pivotal role in extending the frequency range over which both resolved and unresolved \gls{gw} sources can be detected. Jointly, space- and ground-based instruments possess the potential to guide prospective detection efforts regarding the \gls{sgwb}, offering exciting prospects for future \gls{gw} research. \\
Due to its rich phenomenology, the cosmological component within the \gls{sgwb} represents a smoking gun for new physics. However, a direct detection of such poses a substantial difficulty for data analysts: The \gls{sgwb} is predominantly shaped by astrophysical sources, including white dwarf binaries, stellar mass black hole inspirals, and extra-galactic mergers, which augment the diverse range of primordial features in the spectrum.
The task of distinguishing between the instrumental noise, the galactic and the extra-galactic (incl. cosmological) components of the stochastic imprint, without spoiling either of the components, remains one of the major challenges and priorities in the field of \gls{sgwb} measurements \cite{baghi_uncovering_2023, mentasti_probing_2023, bertacca_projection_2020}.
A valuable tool in the complex task of disentangling distinct sources within the \gls{sgwb} is to leverage observer-dependent features that can enhance the detectability against undesirable foreground confusion and instrumental noise. One of the prominent observer-dependent effects within the \gls{sgwb} is characterized by kinematic features arising from the inherent motion of our local group of galaxies \cite{aghanim_planck_2014, Bonvin_2006, cusin_doppler_2022, dallarmi_dipole_2022}. While galactic sources remain unaffected, the (ground- and space-based) detectors' motion with respect to the rest frame of primordial stochastic \gls{gw} sources introduces Doppler anisotropies for primordial contributions to the \gls{sgwb} only. As a result, we anticipate a Doppler-induced increase of power in lower-order multipole patterns of cosmological contributions within the \gls{sgwb}, similar to what has been found in the \gls{cmb} many decades ago. Thus, the analysis of kinematic anisotropies in the SGWB holds considerable promise and has already generated substantial interest within the scientific community \cite{romano_detection_2017, baghi_uncovering_2023, cusin_doppler_2022, bartolo_probing_2022, lisa_cosmology_working_group_maximum_2020}. \\
In this work, we aim to build upon previous investigations and build a diagnostic map-making scheme to be employed primarily (but not exclusively) in the hunt for the extra-galactic sources within \gls{sgwb} data captured by LISA. Our study presents a full time-domain ($4$ years) simulation of GW anisotropic sky via the LISA Simulation Suite (\texttt{LISAGWResponse} \cite{bayle_lisa_2023, bayle_lisa_2022}, \texttt{LISAInstrument} \cite{bayle_unified_2023, lisa_instrument}) and post-processing software (\texttt{PyTDI} \cite{staab_pytdi_2023}) needed to perform the \gls{tdi} combination of the optical measurements and compose the final interferometer observables \cite{hartwig_time-delay_2022}. We utilize a map-making strategy based on a \gls{mcmc} scheme, instead of the commonly used Fisher analysis, to investigate the detectability of kinematic anisotropies in simulated \gls{sgwb} data (see \cite{banagiri_mapping_2021} for similar investigations regarding background from galactic sources). \\
This article is structured as follows:
In section \ref{sec:doppler} we review how Doppler-induced anisotropies manifest in the angular mode decomposition of the energy density spectrum $\Omega_{GW}(f, \mathbf{\hat k})$ of a given source. As for the potential seeds for the tested $\Omega_{GW}(f, \mathbf{\hat k})$, we present three models that are actively discussed in literature and can be associated to physical models of significant phenomenological importance. We elaborate on how anisotropic sky data is processed by LISA, including a description of the response function in section \ref{sec:simulation}. We start by summarizing the data generation procedure in \ref{subsec:data_gen}. Subsequently, based on the simulated interferometer (4-years long) time series data-streams, we present our map-making strategy and the applied \gls{mcmc} scheme in sections \ref{subsec:map} and \ref{subsec:mcmc} respectively, aiming at recovering the injected intensity map. Here, we outline different approaches to recover the lower multipoles from a map reconstructed from LISA data. Finally, we present the result of our simulations for selected \gls{sgwb} features and compare them with respect to their performance with the inclusion of instrumental noise in \ref{sec:analysis}. In section \ref{sec:discussion}, we close our discussion with concluding remarks and an outlook into future work, discussing in particular the improvement of performance of the here presented pipeline in the presences of more complex spectral features.

\section{Doppler-boosted anisotropies as smoking-gun for extra-galactic origin}
\label{sec:doppler}
\begin{figure}
    \centering
    \includegraphics[width=1.\columnwidth]{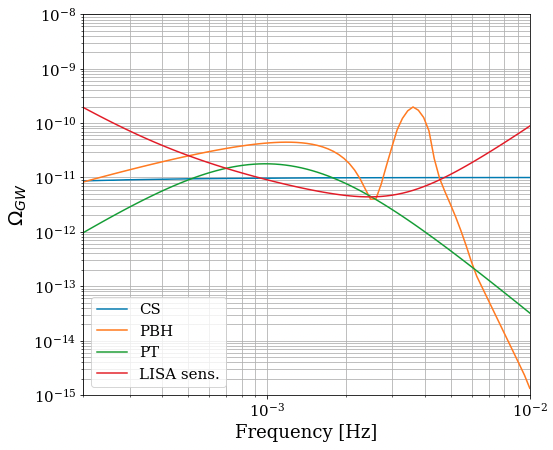}
    \caption{LISA sensitivity curve \cite{Scird} compared to the expected stochastic \gls{gw} spectrum for cosmic strings (blue) modeled as a scale-free contribution at a reference power of $\Omega_{GW}=10^{-11}$, primordial black holes (green) modeled according to \cite{Kohri_2018,PBH}, and first order phase transitions (red) as implemented in the \texttt{PTPlot} package \cite{Caprini_2020}. For the LISA sensitivity curve we use the implementation within the \texttt{PTPlot} package.}
    \label{fig:spectrum}
\end{figure}

We initiate this investigation by briefly outlining the fundamental aspects of the stochastic kinematic anisotropies, basing on \cite{cusin_doppler_2022, bartolo_probing_2022}). In accordance with the introduction, it is established that the \gls{sgwb} account contributions originating both within our galaxy and beyond. The extra-galactic components of the \gls{sgwb} experience a Doppler shift relative to the source frame owing to the motion of our galaxy in relation to the rest frame of the \gls{sgwb}.
Specifically, the behavior of the extra-galactic contributions to the \gls{sgwb} is characterized by a sky-modulation of its apparent (frequency-dependent) energy density, denoted as $\Omega_{GW}(f, \vu{k})$ and where we have made the general frequency and direction dependencies explicit.
To illustrate, we consider two distinct cosmological frames: the first frame, denoted as $\mathcal S'$, is co-moving with the source of the \gls{sgwb} and is referred to as the source frame. The second frame, denoted as $\mathcal S$, is in a state of constant velocity $\mathbf{v}$ relative to $\mathcal S'$ and serves as an observer frame. We make the assumption that the fractional energy density of the \gls{sgwb} in the source frame, denoted as $\Omega'_{GW}(f)$, is perfectly isotropic and solely dependent on frequency $f$. This assumption is valid under the consideration that anisotropies intrinsic to the source are relatively small \footnote{Here, anisotropies arise from factors such as the initial production method of \glspl{sgwb}, Sachs-Wolf (SW) and integrated SW effects, as well as from propagation through a perturbed universe.}. Consequently, without loss of generality, we employ a straightforward Lorentz boost transformation to map the \gls{sgwb} density spectrum in the rest frame $\mathcal S'$, denoted as $\Omega'_{GW}(f)$, to the spectrum in the moving frame, denoted as $\Omega_{GW}(f)$, where $\mathbf{v} = \beta \mathbf{\hat{v}}$ represents the boost velocity \footnote{Throughout, we adopt the convention $c=1$, such that $\beta=|\mathbf{v}|=:v$.}. For the energy density spectrum in particular, the boost mapping $\mathcal S '$ to $\mathcal S$ yields
\begin{align}
    \Omega(f, \mathbf{\hat{k}}) = \left(\frac{f}{f'}\right)^4\Omega(f').
\end{align}
The latter equation is most general for any Lorentz-boost where $0\leq \beta < 1$. Following our assumption of a spectrum being isotropic in the source frame, we can expand this result up to second order on $\beta$, using the relations between $f,f'$ outlined in \cite{Doppler_effort_IV}, i.e. 
\begin{align}\label{equ:-1}
    \frac{f}{f'} = \mathcal{D} = \frac{\sqrt{1-\beta^2}}{1-\beta\  \vu{k}\cdot\vu{v}}
\end{align}
such that
\begin{align}\label{equ:0}
    \Omega_{GW}(f, \mathbf{\hat{k}}) = \mathcal{D}^4 \ \Omega_{GW}'\left( \mathcal{D}^{-1} f\right)
\end{align}
\begin{align}\label{equ:1}
    \Omega_{GW}(f,\mathbf{\hat{k}})=\Omega'_{GW}(f)\bigg(&1+M(f)+\mathbf{\hat{k}\cdot\hat{v}}D(f)
    \notag \\&+
    \left[\left(\mathbf{\hat{k}\cdot\hat{v}}^2-\frac{1}{3}\right)Q(f)\right]\bigg).
\end{align}
Here, the functions $M(f), D(f), Q(f)$ correspond to the monopole, dipole, and quadrupole contributions respectively. These functions are given by 
\begin{align}\label{equ:M}
    M(f)&=\frac{\beta^2}{6}\left(8+n_{\Omega}(n_{\Omega}-6)+\alpha_{\Omega}\right),
\end{align}
\begin{align}\label{equ:D}
    D(f)&=\beta(4-n_\Omega),
\end{align}
\begin{align}\label{equ:Q}
    Q(f)&=\beta^2\left(10-\frac{9n_\Omega}{2}+\frac{n^2_\Omega}{2}+\frac{\alpha_\Omega}{2}\right),
\end{align}
and
\begin{align}\label{equ:alpha_n}
    n_\Omega(f)=\frac{\ddd \ln(\Omega'_{GW}(f))}{\ddd \ln{f}},&&\alpha_{\Omega}(f)=\frac{\ddd n_{\Omega}(f)}{\ddd \ln{f}}.
\end{align}
It is crucial to acknowledge that the functions $M(f)$, $D(f)$, and $Q(f)$ introduced now are entirely model-dependent and encapsulate distinctive features corresponding to specific types of sources contributing to the extra-galactic segment of the \gls{sgwb}. An expansion in $\mathbf{\hat{k}}\cdot \mathbf{\hat{v}}$ emerges as a consequence of the expansion in $\beta$. It is essential to highlight that $M(f)$, $D(f)$, and $Q(f)$ may acquire additional contributions beyond those specified in Equations \eqref{equ:M}-\eqref{equ:Q} if intrinsic anisotropies are present in the spectrum of the source frame. As a result, Equation \eqref{equ:1} when used in conjunction with $M(f), D(f)$, and $Q(f)$ as defined, is only valid when anisotropies are absent from the source frame.
As computationally evident, the aforementioned functions determine the harmonic expansion coefficients $a_{lm}$ up to constant pre-factors. These coefficients play a fundamental role in shaping the angular power spectrum $C_{l}^{GW}$. By rewriting \eqref{equ:1} as
\begin{align}
    \Omega_{GW}(f,\mathbf{\hat{k}})=\Omega'_{GW}(f)\left(1+\delta_{GW}^{\text{kin}}(f,\theta,\phi)\right),
\end{align}
 where $\delta_{GW}^{\text{kin}}$ now describes the anisotropic part of the measured energy density $\Omega_{GW}$ in the observer frame $\mathcal S$, we can exploit that $\delta_{GW}^{\text{kin}}$ has a smooth angular dependence, and expand as
 \begin{align}
     \delta_{GW}^{\text{kin}}(f,\mathbf{\hat{k}})=\sum_\ell\sum_m\delta_{GW,\ell m}^{\text{kin}}(f)Y_{\ell m}(\mathbf{\hat{k}}).
 \end{align}
 The angular power of the anisotropies is then given as an ensemble average
 \begin{align}\label{equ:3}
\langle\delta_{GW,\ell m}^{\text{kin}},\delta_{GW,\ell'm'}^{\text{kin}}\rangle=:C_\ell^{GW}(f)\delta_{\ell \ell'}\delta_{mm'}.
 \end{align}
 The coefficient $C_\ell^{GW}$ can as well be derived by simply inserting the anisotropies $\delta_{GW}^{\text{kin}}(f,\mathbf{\hat{k}})$ in the latter equation. Then, by orthogonality of the spherical harmonics we find\footnote{Note that here we use the convention $\int_0^\pi\dd \theta\int_{\varphi=0}^{2\pi}Y_{\ell m} \, Y_{\ell'm'}^*d\Omega = \frac{4 \pi}{(2 \ell + 1)} \delta_{\ell\ell'}\, \delta_{mm'}$.}
 \begin{align}
     &\frac{1}{4\pi} \int_{\Omega}\left|\delta_{GW}^{\text{kin}}\right|^2\ddd \Omega\notag\\
     =&\sum_\ell\frac{1}{2\ell+1}\sum_m\left|\delta_{GW,\ell m}^{\text{kin}}\right|^2=:\sum_\ell C_\ell^{GW},
 \end{align}
 therefore,
 \begin{align}\label{equ:4}
     C_\ell^{GW}=\frac{1}{2\ell+1}\sum_{m=-\ell}^{\ell}\left|\delta_{GW,\ell m}^{\text{kin}}(f)\right|^2.
 \end{align}
If we now combine this result with Equation \eqref{equ:1}, we can assign each mode for $\delta_{GW,lm}^{\text{kin}}$ to the corresponding kinematic mode, that is\footnote{We here choose the frame where $\mathbf{\hat{k}}\cdot \mathbf{\hat{v}\approx \cos \theta}$. In subsequent sections, we transition to the International Celestial Reference System (ICRS) frame and concentrate on \eqref{equ:0}, albeit expressing it in terms of monopole, dipole, and quadrupole language utilizing the expansion coefficients $M(f)$, $D(f)$, and $Q(f)$.}
\begin{align}
    \delta_{GW,00}^{\text{kin}}(f)=2\sqrt{\pi}M(f),&&\delta_{GW,10}^{\text{kin}}(f)=2\sqrt{\frac{\pi}{3}}D(f) ,\notag&&\\ \delta_{GW,20}^{\text{kin}}(f)=\frac{4}{3}\sqrt{\frac{\pi}{5}}Q(f).
\end{align}
It follows trivially that
\begin{align}
    C_0^{GW}\sim|M(f)|^2, && C_1^{GW}\sim|D(f)|^2,\notag\\
    C_2^{GW}\sim|Q(f)|^2.
\end{align}
It is crucial to underscore that the outlined decomposition offers distinct advantages. For instance, Equations \eqref{equ:M}-\eqref{equ:Q} are determined by the gradient of the spectrum and its derivatives. It is easy to show that, in cases of a rapid change in the frequency spectrum, such as peaks or discontinuities, the individual components $M(f)$, $D(f)$, and $Q(f)$ experience an amplification in their amplitude, driven by the $n_\Omega$ and $\alpha_\Omega$. In some cases, we find that this amplification effect leads to the kinematic quadrupole dominating the kinematic dipole in the signal. The direct implications of such phenomena are discussed in section \ref{sec:discussion}.\\
Choosing a particular instance for the spectrum $\Omega_{GW}$, we leverage three principal examples of early universe cosmology that manifest in the \gls{sgwb}. The significance of these signals arises from their anticipated amplitude and frequency range, aligning with the sensitivity of LISA \cite{Scird}, as illustrated in Figure \ref{fig:spectrum}. The main spectrum under consideration exhibits an approximate scale-free nature. In our analysis, it resembles the energy density spectrum of \gls{cs}, which is anticipated to manifest as a nearly slope-free spectrum \cite{PhysRevLett.98.111101, CHANG2020100604, PhysRevD.98.063509} within the pertinent frequency range of LISA. \footnote{For a given model $i$, the relevant frequency regime is defined as the intersection between the energy density $\Omega_{GW}^i$ and the region above the LISA sensitivity curve, as depicted in Figure \ref{fig:spectrum}.} 
Despite its simplicity, the analysis of \gls{cs} features in the \gls{sgwb} is intriguing, as \gls{cs} serve as potent probes of physics beyond the Standard Model in the early universe and hold relevance in the context of string theory \cite{CS_I, CS_II}.\\
The second spectrum under consideration is sourced by a first-order phase transition (PT) in the early universe and anticipated to follow a broken-power-law \cite{caprini_detecting_2020, Caprini_2020} as depicted in Figure \ref{fig:spectrum}\footnote{Note that the exact shapes of the spectra, i.e. peak frequency $f_{\text{peak}}$ or slope does not play a significant role in this analysis. Solely the fact that non-trivial $n_\Omega$ and $\alpha_\Omega$ are achieved matters here.}.
Phase transitions offer valuable phenomenological insights, notably marking scales of symmetry breaking occurring in the early universe. \\
Finally, our interest extends to the spectrum resulting from primordial black holes (PBH), as described in \cite{Kohri_2018,PBH}. However, it is acknowledged at this point that the literature lacks consensus on the exact shape in frequency space of PBH contributions to the \gls{sgwb}. Regardless of the exact shape of the spectrum, the phenomenology associated with PBHs is extensive. Notably, models proposing Dark Matter compositions involving PBHs, either partially or entirely, have recently garnered significant interest within the scientific community \cite{Carr_2016, Carr_2020}. \\
In the main analysis, i.e. section \ref{sec:analysis}, we focus on the simplest featureless model, the CS-like signal, and investigate its kinematic anisotropies through a realistic full-time domain simulation. The impact of more feature-full spectra such as the ones resulting from PT and PBH on the here presented analysis pipeline is discussed \ref{sec:discussion}.\\

\section{Anisotropic GW stochastic sky simulation and instrument response}
\label{sec:simulation}
This paper puts forth an end-to-end, time-domain simulation approach to generate realistic LISA data, coupled with a specialized analysis pipeline crafted for the identification of $l=0,1,2$ (kinematic) anisotropies within a \gls{sgwb} signal. 
To establish context, we describe the interplay between the detector's response and the \gls{sgwb} signal in the following.\\
The \gls{sgwb} signal can be modeled as a random, sky-dependent strain time-series $h_{P}(t, \vu{k})$ of polarization $P$, following Gaussian statistics \cite{lisa_cosmology_working_group_maximum_2020}. This time-series is fully characterized by the second-order moments of the Fourier transforms of the strain, denoted as $\langle {h}_{P}(f, \vu{k}) {h}^*_{P'}(f, \vu{k}) \rangle$. This expression defines the cross-power spectra $S_{PP'}(f,\hat{\vb{k}})$ of the stochastic process $h_{P}(t, \hat{\vb{k}})$. Under the assumption of statistical homogeneity, $\langle {h}_{P}(f, \hat{\vb{k}}) {h}^*_{P'}(f, \hat{\vb{k}}) \rangle$ corresponds to 
\begin{align}
S_{PP'}(f,\vu{k}) & =
\begin{bmatrix}
    \langle {h}_{+}(f, \vu{k}) {h}^*_{+}(f, \vu{k}) \rangle &  \langle {h}_{+}(f, \vu{k}) {h}^*_{\times}(f, \vu{k}) \rangle \\
    \langle {h}_{\times}(f, \vu{k}) {h}^*_{+}(f, \vu{k}) \rangle & \langle {h}_{\times}(f, \vu{k}) {h}^*_{\times}(f, \vu{k}) \rangle
\end{bmatrix}\notag
\\
& = \frac{1}{2} \delta^2 (\vu{k}-\vu{k}')\delta(f,f')
\begin{bmatrix}
    I + Q &&  U + i V  \\
    U - i V &&  I - Q  \\
\end{bmatrix},\notag
\end{align}
where one can introduce the Stokes parameters $I, Q, U$ and $V$, well-known in CMB physics and encoding intensity, linear polarization and circular polarization respectively. As in \cite{lisa_cosmology_working_group_maximum_2020}, this analysis is restricted to the intensity $I(f, \vu{k})$ with 
\begin{equation}\label{equ:I_def}
    I(f, \hat{\vb{k}}) = \langle {h}_{+}(f, \hat{\vb{k}}) {h}^*_{+}(f, \hat{\vb{k}}) \rangle + \langle {h}_{\times}(f, \hat{\vb{k}}) {h}^*_{\times}(f, \hat{\vb{k}}) \rangle
\end{equation}
which can be related to the normalized logarithmic energy density $\Omega_{GW}(f,\vu{k})$ as \cite{allen_detection_1997}
\begin{align}\label{equ:energydensity}
    \Omega_{GW}(f,\vu{k}) = \frac{32\pi^3 f^3}{3H_0^2}I(f,\vu{k}),
\end{align}
where $H_0$ is the Hubble constant.
Note here that the direction dependence of the latter two equations is commonly dropped by the assumption that
\begin{align}
    \Omega_{GW}(f, \vu{k}) = \Omega_{GW}(f) \mathcal{E}(\vu{k}).
\end{align}
The first factor on the right-hand side of Equation \eqref{equ:energydensity} remains applicable, while the second factor encapsulates the angular distribution of the background. We chose a normalization such that 
\begin{align}
    \int_{\mathcal{S}^2}\dd^2 \vu{k}\, \mathcal{E}(\vu{k})=1.
\end{align}
With the stochastic strain ${h}_P$ and the associated intensity $I(f, \vu{k})$ at hand, we can now characterize the incoming signal as a time-frequency series. Following Equation (12) in \cite{lisa_cosmology_working_group_maximum_2020}, we define the signal component of the time stream $s^\tau_I$ measured by a single \gls{tdi} channel $C$ as a Fourier expansion between $t$ and $t+\Delta t$, such that it reads
\begin{align}\label{equ:defresponce}
s^\tau_C(f)=\sum_P\int_{S^2}\dd\vu{k}\,R^{\tau,P}_C(f,\vu{k}) h_P(f,\vu{k}).
\end{align}
The superscript $\tau$ indicates that $s^\tau_C(f)$ is a potentially time-varying frequency series. In this expression, $R^P_C$ represents the LISA response function, which depends on the selected channel $C \in [X, Y, Z]$ and the polarization $P \in [+, \times]$ of the strain. The elements $R^P_C$ and the concept of channels will be discussed in the next subsection. Based on the definitions above, it holds that 
\begin{align}
    & \langle{h_P(f,\vu{k}),h^*_{P'}(f',\vu{k}')\rangle}\notag \\
    &=\frac{1}{2}\frac{1}{4\pi}\delta_{f,f'}\delta^2(\vu{k},\vu{k}')\delta_{P,P'}S_{PP'}(f,\vu{k}).
\end{align}
Under the Gaussian assumption, the power spectrum $S_P(f,\vu{k})$ becomes the primary measurable quantity. The factor of $4\pi$ in the denominator arises from an integral over the unit sphere. While intensity is a quantity defined per pixel, the power spectrum density is an integral over the entire sky; hence, they differ by a factor of $4\pi$. We can further assume that the \gls{sgwb} is not polarized, so that $S_+(f,\vu{k})=S_\times(f,\vu{k})=\frac{1}{2}S_{GW}(f,\vu{k})$ where the latter can be conveniently characterized by $\Omega_{GW}(f,\vu{k})$ via \cite{caprini_cosmological_2018}
\begin{align}\label{equ:sim_two}
    S_{GW}(f,\vu{k})=\frac{3H_0^2}{4\pi^2f^3}\Omega_{GW}(f,\vu{k}).
\end{align}
Note at this point the similarities between Equations \eqref{equ:energydensity} and \eqref{equ:sim_two}. 

\subsection{LISA response function}
\begin{figure}[t!]
    \centering
    \includegraphics[width=0.8\columnwidth]{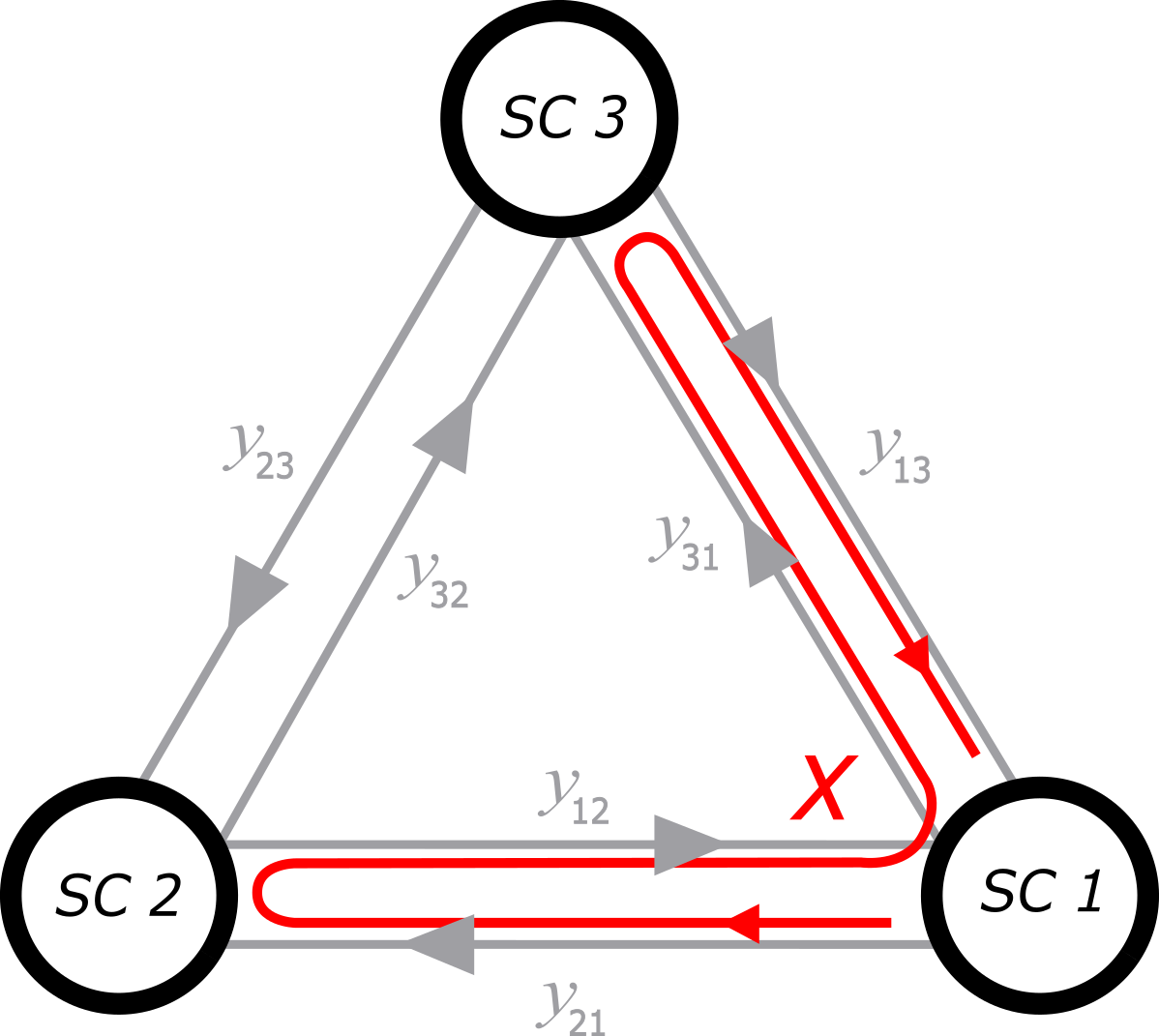}
    \caption{Illustration of three LISA spacecrafts in triangular formation, connected via six links (gray). Displayed in red is the 1.5 TDI $X$ channel as a linear combination of links. }
    \label{fig:LISA_TDI}
\end{figure}

The three spacecraft composing the LISA instrument contain six distinct links, illustrated in Figure \ref{fig:LISA_TDI}, each of which is deformed in a time-dependent manner by incoming gravitational radiation. Due to the linearity of the response function, the overall response of link $ij\in\{12,21,13,31,23,32\}$ is given by the sum of individual responses to a source allocated in pixel $p$,
\begin{align}\label{equ:indiv_resp}
    \mathbf{\tilde{y}}^{\tau}(f) = y^\tau_{ij}(f) = \sum_p y_{ij,p}^\tau(f).
\end{align}
Here again, we denoted time dependence via the superscript $\tau$. The expression of the total link response in terms of response per pixel is required for numerical modeling, and is anyhow motivated by the limited resolution of the instrument. For every time step $\Delta t$, $y_{ij,p}$ can be understood as the frequency shift induced on the laser beam along the link $ij$ by the gravitational strain originating in the pixel $p$. To obtain an explicit expression for $y_{ij,p}$, we hence have to project the strain from point source $p$ onto the unit vector pointing along the link $ij$. Under the approximation of immobile spacecrafts during the light propagation along a single link, one finds \cite{baghi_uncovering_2023}
\begin{widetext}
\begin{align}
    y_{ij,p}(t) \approx \frac{1}{2\left(1-\hat{\vb{k}}_p\cdot \hat{\vb{n}}_{ij}(t)\right)}\left[H_{ij,p}\left(t-\frac{L_{ij}(t)}{c}-\frac{\hat{\vb{k}}_p\cdot \vb{n}_j(t)}{c}\right)-H_{ij,p}\left(t-\frac{\hat{\vb{k}}_p\cdot \vb{n}_i(t)}{c}\right)\right].
\end{align}
\end{widetext}
In the latter, $L_{ij}(t)$ is the time-dependent separation between two spacecrafts defined via their positions $\vb{x}_i=\vb{x}_j + L_{ij}\hat{\vb{n}}_{ij}$ and $\hat{\vb{n}}_{ij}$ is the unit link vector. In the case of $c=1$, $L_{ij}(t)$ corresponds to the delay time along the link $ij$ at reception time $t$. The vector $\vu{k}_p$ represents the wave vector of the \glspl{gw} originating from pixel $p$, but also understood as the direction in the sky of the pixel $p$ (in the plane wave approximation). The projection function $H_{ij,p}$ is given by
\begin{align}
    H_{ij,p}(t) & = h_+(t,\vb{\hat{k}}_p)\ \xi_+(\hat{\vb{u}}_p,\hat{\vb{v}}_p,\hat{\vb{n}}_{ij}) \\
    & + h_\times(t,\vb{\hat{k}}_p)\ \xi_\times(\hat{\vb{u}}_p,\hat{\vb{v}}_p,\hat{\vb{n}}_{ij})\notag,
\end{align}
where the functions $\xi_{+}$ and $\xi_{\times}$ are the \textit{antenna pattern functions} and $\hat{\vb{u}}_p,\hat{\vb{v}}_p$ the polarization vectors associated to the propagation vector $\hat{\vb{k}}_p$. For details, see appendix A in \cite{baghi_uncovering_2023} and references therein.
\\
To compute TDI observables, LISA combines the six link-signals resulting in three correlated channels commonly referred to as $X,Y,Z$, illustrated in Figure \ref{fig:LISA_TDI}. The explicit linear combination of links leading to the designated channels can be encapsulated in one matrix, $\mathbf M_{TDI}$ \cite{baghi_uncovering_2023}. For instance, the second generation \gls{tdi} $X_2$ channel as it is sketched in Figure \ref{fig:LISA_TDI} can be constructed using links $y_{12},y_{21},y_{13},y_{31}$, 
\begin{align} X_2 =&\, X_1 + \vb{D}_{1321}y_{12}+\vb{D}_{131212}y_{21}+\vb{D}_{1312121}y_{13}\notag\\&+\vb{D}_{13121213}y_{31}- (\vb{D}_{12131}y_{13}+\vb{D}_{121313}y_{31}\notag\\&+\vb{D}_{1213131}y_{12}+\vb{D}_{12131312}y_{21}),\label{equ:X2}
\end{align}
with
\begin{align}
    X_1 =& \,y_{13} + \vb{D}_{13}y_{31} + \vb{D}_{131}y_{12}+ \vb{D}_{1312}y_{21}\notag\\
    &- ( y_{12} + \vb{D}_{12}y_{21}+\vb{D}_{121}y_{13}+\vb{D}_{1213}y_{31}).\label{equ:X1}
\end{align}
The delay operator for a single link is defined as 
\begin{align}
    \vb{D}_{ij}x(t)= x(t-L_{ij}(t)),
\end{align}
and can be accumulated forming the chained delay operator, $\vb{D}_{i_1,i_2,..}$, which, applied on a function $f(t)$, induces a chained delay in the reception, i.e.
\begin{align}
    \vb{D}_{i_1,i_2,..}f(t) = f\left(t- \sum_{l=1}^{n-1} L_{i_l,i_l+1}(t)\right).
\label{eq: total_delay}
\end{align}
Analogous construction can be done for the $Y,Z$ channels by permuting $ij$ in \eqref{equ:X1} and \eqref{equ:X2} correspondingly.

\subsection{Analytical frequency domain response model and angular dependence}

In the frequency domain, delay operators in the Equation \myhyperref{eq: total_delay} are expressed simply as phase operators. Using the analytical expression derived in Equation (B.5) of \cite{baghi_uncovering_2023}, we write in Equation \myhyperref{eq: kernel} the single-link frequency series contributed from the unit sky-direction $\vu{k}$ as the projection of the two polarizations to the frequency-domain response $2\times 6$ (two polarizations, six links) kernel $G_{ij,P}^\tau(f,\vu{k})$.
\begin{equation}
\mathbf{\tilde{y}}^{\tau}_{p}(f) = \sum_{P=(+,\times)} G^\tau_{ij,P}(f, \vu{k}_p)\ h_{P}^{\tau}(f,\vu{k}_p)
\label{eq: kernel}
\end{equation}
We will use the approximation that the transfer functions $G^\tau_{ij,P}(f, \vu{k}_p)$ are stationary within the time segment labeled by $\tau$, which implies time windows short enough relatively to the LISA orbital timescale ($\ll 1 \text{year}$). Then, one can structure the delay phase operators in a matrix form $\mathbf M^\tau_{TDI}(f)$ and whose time dependence ($\tau$ upper script) is induced by the annual orbital breathing of LISA constellation \cite{baghi_uncovering_2023}. The measured data within a discrete time and frequency interval, $\mathbf{\tilde{d}}^{\tau}(f) \equiv (\tilde X,\tilde Y,\tilde Z)^T$, can be expressed as a vector of \gls{tdi} channels:
\begin{align}\label{equ:data_vecI}
    \mathbf{\tilde{d}}^{\tau}(f,\vu{k}) = \mathbf M_{TDI}^\tau (f)\ \mathbf{\tilde{y}}^{\tau}(f,\vu{k}),
\end{align}
with $\mathbf{\tilde{y}} = (\tilde{y}_{12},\tilde{y}_{23},\tilde{y}_{31},\tilde{y}_{13},\tilde{y}_{32},\tilde{y}_{21})^T$ and where we made the time dependence an explicit argument of each tensor. \\
Using definition \eqref{equ:defresponce} and the explicit formula for the individual link response, one can construct a three-vector (see appendix B in \cite{baghi_uncovering_2023})
\begin{align}
    \vb{R}_{P}^\tau(f,\vu{k}) = \mathbf M^\tau_{TDI}(f) \ G^\tau_{ij,P}( f, \vu{k})\ \mathbf M^\tau_{TDI}(f)^{\dag},\notag
\end{align}
that contains the individual response components for each 1.5 TDI channel $X,Y,Z$. The subscript indicates the polarization dependence induced by the strain appearing in \eqref{equ:defresponce}\footnote{Compare also \cite{lisa_cosmology_working_group_maximum_2020}.}. The linear response TDI vectors $\vb{R}^P=(R_X,R_Y,R_Z)^P$ can be in turn merged into the quadratic response $\vb{A}^\tau$ ($N_f \times N_\tau \times 3 \times 3 \times N_{\text{pix}}$):
\begin{align}
\vb{A}^\tau(f,\vu{k})=\vb{R}^+\otimes \vb{R}^{+*}+\vb{R}^\times \otimes \vb{R}^{\times*}, 
\end{align}
such that the covariance matrix for a measured intensity $\Tilde{I}$ and noise matrix $\vb{N}$ reads
\begin{align}
    \langle\vb{S}^\tau(f)\rangle = \vb{C}^\tau_f \approx \sum_p \vb{A}_p\Tilde{I}^p + \vb{N},
\label{eq: covariance_raw}
\end{align}
defined based on the quadratic strain tensor
\begin{align}
    \langle\vb{S}^\tau(f)\rangle &= \int_{S^2}\dd\vu{k}\, \vb{s}^\tau(f) \otimes \left(\vb{s}^\tau(f)\right)^* \notag \\ &= \int_{S^2} \dd\vu{k}\,\vb{A}^\tau(f, \vu{k}) I(f,\vu{k}).
    \label{equ: quadratic strain}
\end{align}
In the latter, we use a vector comprised of components \eqref{equ:defresponce} to construct a matrix in channel-space. In particular, Equation \eqref{equ: quadratic strain} and hence also the covariance matrix \eqref{eq: covariance_raw} itself are frequency- and direction-dependent time series of $3\times 3$ matrices where the entries correspond to channels. Similarly, a TDI data vector recorded by LISA contains three components reflecting the three channels as mentioned above. Equation \eqref{equ:data_vecI} can be decomposed into response, signal and noise via 
\begin{align}
    \vb{\tilde{d}} = \vb{R}h + \vb{n},
\label{eq: data - linear}
\end{align}
where the TDI noise $\vb{n}$ is assumed to be Gaussian with zero mean and a covariance $\vb{N}_{d} = \vb{n}\otimes \vb{n}$. The data components are assumed to be Gaussian with covariance given by Equation \eqref{eq: covariance_raw}. Note that $\vb{\tilde{d}}$ is again time and frequency dependent.
However, note that the individual link response \eqref{equ:indiv_resp} is sensitive to the pixels in the sky and hence the direction. Consequently, so is $\vb{R}^P$. It is worth mentioning that there are analytical approaches for describing $\vb{R}^P$ for the low frequency regime of LISA's sensitivity band, such as the one outlined in \cite{lisa_cosmology_working_group_maximum_2020}. However, we aim to harness the power of numerical tools to provide a more realistic detection scenario and hence rely on the numerical implementations. More details are given in the subsequent subsection.\\
In Equation \eqref{equ: quadratic strain}  the product $\vb{A}\Tilde{I}$ is integrated over the full sky. However, with a realistic finite sky resolution one preferably rewrites the integral as the sum over pixels, i.e. $\vb{A}_p\Tilde{I}^p$. Converting angular- to pixel-dependence results in the response matrix being a $3$-dimensional $3\times 3\times N_\text{pix}$ frequency- and time-dependent matrix. Here, $N_\text{pix}$ stands for the number of pixels. The covariance matrix, due to the full sky integration in \eqref{eq: covariance_raw}, is direction-independent.

\subsection{Spherical harmonics representation}
\label{subsection: angular response}

For the purpose of this work, tracing over pixels is the preferred measure of full-sky summation, simply due to the numerical nature of these analyses. Nevertheless, on an analytical level, we can exploit the continues basis of spherical harmonics to decompose the covariance matrix into individual modes. This results in analog decomposition of the intensity $I$ and consequently the power spectrum $\Omega_{GW}$ similar to Equation \eqref{equ:1}. Note however that in contrast to \eqref{equ:1} the decomposition in modes does not require any assumptions about the (kinematic) origin of the signal. Hence, this approach stays more agnostic with respect to the underlying model.\\
Although $\vb{A}$ and $I$ depend on the direction in the sky, in Equation \eqref{equ: quadratic strain} the Mollweide projection provides a bijective map transforming $\vb{A(\hat{\vb{k}})}\rightarrow\vb{A}^p$ and vice versa. As a continuous map can be decomposed into spherical harmonics, the Mollweide map enables us at any point of our analysis to write the covariance matrix in terms of modes rather than pixels. In practice the mapping can be achieved as follows: Writing the sum over pixels explicitly, we can replace the latter by the approximation
\begin{align}\label{equ:pix_vs_smooth}
    \sum_p \vb{A}_p\Tilde{I}^p \approx \frac{1}{\Delta_{\text{pixel}}} \int_{S^2} \dd \mathbf{\vu{k}}\,\vb{A}(\mathbf{\vu{k}})\Tilde{I}(\mathbf{\vu{k}}).
\end{align}
Here, $\Delta_{\text{pixel}}$ corresponds to the area per pixel \footnote{Note that for most commonly used mappings of the Riemann sphere $\mathcal S^2$ this area per pixel is not constant, however we will refer to suitable python packages taking care of this transformation.}. Frequency and time dependence of both response matrix and intensity are omitted in the latter. Naturally, the approximation improves with the number of pixels. The latter equation can be simplified further by rewriting $\vb{A}$ and $I$ as
\begin{align}\label{equ:mode_decomp}
    \vb{A}(\vu{k})=\sum_{\ell,m}a_{\ell,m}Y_{\ell,m}(\vu{k}),&&I(\vu{k})=\sum_{\ell,m}i_{\ell,m}Y_{\ell,m}(\vu{k})
\end{align}
where
\begin{align}
    & \langle i_{\ell m},i^{*}_{\ell'm'} \rangle =: C_\ell^{GW}\delta_{ll'}\delta_{mm'}, \notag \\
    & \langle a_{\ell m},a^{*}_{\ell'm'} \rangle =: \vb{A}_\ell\delta_{\ell \ell'}\delta_{mm'}
\end{align}
and 
\begin{align}
    C_\ell^{GW}= \frac{1}{2\ell+1}\sum_{m=-\ell}^\ell\left|\int_S^2\frac{\dd\vu{k}}{4\pi}Y_{\ell,m}(\vu{k})I(\vu{k})\right|^2,
\end{align}
\begin{align}
\vb{A}_\ell= \frac{1}{2\ell+1}\sum_{m=-\ell}^\ell\left|\int_S^2\frac{\dd\vu{k}}{4\pi}Y_{\ell,m}(\vu{k})\vb{A}(\vu{k})\right|^2,
\end{align}
so that 
\begin{align}
    \vb{A}_pI^p\rightarrow \int_{S^2}\dd\vu{k}\sum_{\ell',m'}\sum_{\ell,m}a_{\ell m}i_{\ell'm'}Y_{\ell',m'}(\vu{k})Y_{\ell,m}(\vu{k}).
\end{align}
Note though that, technically, we neglected the factor of the pixel area here and mathematical equality in the latter holds only in the limit of large pixel number. In either case, the right-hand side can be simplified by using that $Y_{\ell,-m}=(-1)^mY_{\ell,m}^*(\vu{k})$ and $i_{\ell,-m}=(-1)^mi^*_{\ell m}$\footnote{This relation only holds for the mode components of $I(\mathbf{\hat{n}})$ as the intensity is a real quantity, as opposed to the response.}, and the orthogonality relation of the spherical harmonics. Then we find
\begin{align}
&\int_{S^2}\dd\vu{k}\sum_{\ell',m'}\sum_{\ell,m}a_{\ell m}i_{\ell'm'}Y_{\ell',m'}(\vu{k})Y_{\ell,m}(\vu{k})\notag\\
&=\int_{S^2}\dd\vu{k}\sum_{\ell,m}\sum_{\ell',m'}a_{\ell m}i^{*}_{\ell',-m'}Y_{\ell m}(\vu{k})Y^*_{\ell',-m'}(\vu{k})\\
    &=\int_{S^2}\dd\vu{k}\sum_{l,m}\sum_{\ell',m'}a_{\ell m}i^{*}_{\ell'm'}Y_{\ell m}(\vu{k})Y^*_{\ell',m'}(\vu{k})\notag\\
    &=\sum_{\ell,m}a_{\ell m}i^{*}_{\ell m},
\label{bla^3}
\end{align}
and thus
\begin{align}\label{equ:1.234}
    \vb{A}_pI^p \approx \frac{1}{\Delta_{\text{pixel}}}\sum_{\ell<3}\sum_{m=-\ell}^{\ell}a_{\ell m}i^{*}_{\ell m}.
\end{align}
We sum only over $\ell=0,1,2$, and the remaining modes are truncated, assuming that the signal intensity $I$ is significant only up to $\ell=2$, i.e., up to quadrupole contributions, analogous to \eqref{equ:1}. It is noted at this point that this is equivalent to setting $i_{\ell m} = 0$ for $\ell >2$.\\ 

\subsection{Data generation}
\label{subsec:data_gen}
\begin{figure*}
\centering
\frame{\includegraphics[width=\linewidth, trim={0.0cm 0.8cm 0.0cm 2.5cm}, clip]{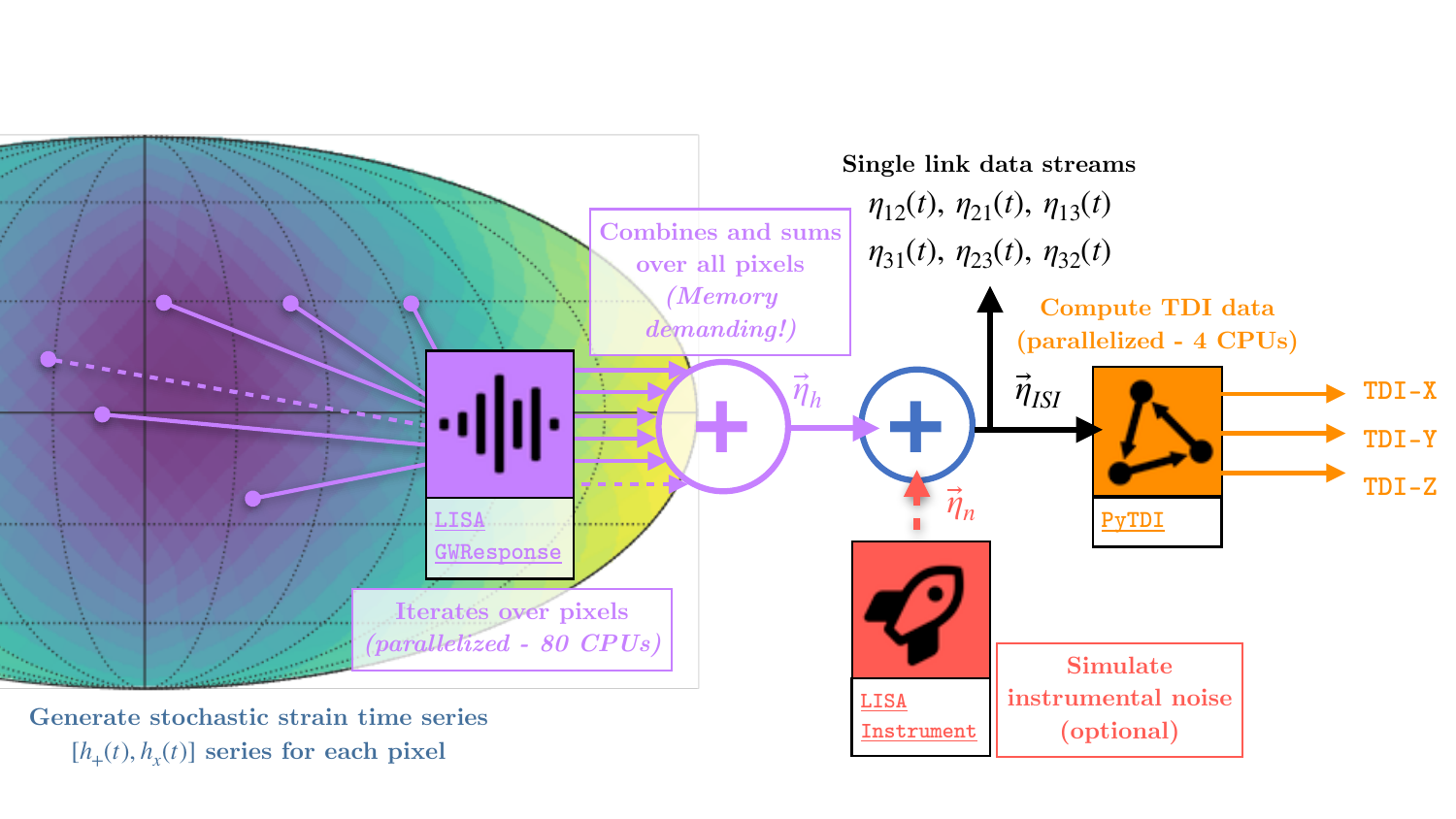}}
\caption{\gls{e2e} simulation flow of \gls{lisa} response to an anisotropic, stochastic \gls{gw} sky.}
\label{fig: sim flow}
\end{figure*}
Our proposal involves simulating the response of the LISA detector to a Doppler-boosted, anisotropic \gls{sgwb} sky. Specifically, we aim to generate a synthetic dataset spanning 4 years, as outlined in Equation \eqref{eq: data - linear}, with covariance given by Equation \eqref{eq: covariance_raw}. For the data generation, we will focus solely on the simple case of an \gls{sgwb} with a flat energy density spectrum, such as potential CS signals in the LISA band \cite{caprini_cosmological_2018}.\\
For the generation of data, we employ an angular discretization of the sky into finite-size pixels using the python package \href{https://healpy.readthedocs.io/en/latest/}{\texttt{healpy}}. The latter is extensively used throughout our analysis, providing a useful tool for map-mapping and the conversion from the pixel to mode domain. Each pixel of the generated $N_{\text{pix}} = 12 \times 2^{N_{\text{side}}}$ pixels sky map is simulated as an individual, independent stochastic strain time series of \gls{psd}, $S_h$, as given in Equation \eqref{equ:sim_two}. We emphasize here that, initially, we do not include instrumental noise or confusion foregrounds in the data. Hence, the first step of the analysis is solely dedicated to analyzing \gls{lisa}'s capabilities in dealing with faint signals of stochastic nature. Hereby, the complexity to be overcome resides in the randomness of the time series for each pixel in the sky (see Figure \ref{fig: sim flow}). \\
The \gls{lisa} single-link responses to each individual pixel is computed using, as in \cite{baghi_uncovering_2023}, the \href{https://pypi.org/project/lisagwresponse/}{\texttt{LISAGWResponse}} software \cite{bayle_lisa_2023}, part of the LISA consortium simulation suite, which provides a time-domain projection of $h_+$ and $h_x$ polarization time series on the antenna response with minimal approximations \footnote{Such include a static spacecraft within arm light-travel timescale as well as first order GW propagation time expansion.}. In addition, a simplified orbital set-up is used, which considers an equilateral, equal arm configuration \cite{bayle_lisa_2022}. The linearity of the response function allows computing the overall response to an anisotropic \gls{gw} sky as a net sum of the $N_{\text{pix}}$ single pixel source responses. This projection is hence typically parallelized and distributed over up to $>80$ CPUs, selecting an adequate trade-off between CPUs and Memory usage (i.e. the more CPU-distributed, the more RAM demanding). For an $N_{\text{side}} = 32$ sky resolution, $4$-years long dataset sampled at $0.2 \si{\hertz}$, the data generation deployed on a computing node uses up to 128 CPUs and $3 \si{\tera\byte}$ RAM over $15$ hours or runtime. Particular attention should be set on the statistical independence across the stochastic pixel sources when parallel computation is performed. Each individual process initializes the random generators with a specific {\it local seed} while a {\it global seed} reshuffling the overall sky generation ensures statistical independence between simulation runs.\\
The single-link simulated data are subsequently processed using the LISA consortium \texttt{PyTDI} software \cite{staab_pytdi_2023} to generate the second-generation Time-Delay Interferometry (\gls{tdi}) time-domain Michelson-like interferometer data streams $X_{2}(t)$, $Y_{2}(t)$, and $Z_{2}(t)$. Optionally, we introduce instrumental noise at the single-link level using the LISA simulation suite software \href{https://pypi.org/project/lisainstrument/}{\texttt{LISAInstrument}}, enabling only the secondary noises for simplicity and efficiency, as the \texttt{PyTDI} software downstream is dedicated and designed to suppress the primary noises. Figure \ref{fig: sim flow} summarizes the end-to-end simulation flow.\\
It is crucial to highlight that the clock frame with respect to which phase measurements are recorded and stamped has a critical impact, when beat note data streams are combined to produce the Time-Delay Interferometry (\gls{tdi}) time-series. While using local spacecraft reference time to compose the $X$, $Y$, and $Z$ data streams is physically correct, as they are all reducible to observations made on a single spacecraft located at the node of the Michelson-like interferometer \cite{hartwig_time-delay_2022}, issues arise when combining these time-series ($X$, $Y$, and $Z$) based on different time frames. They need to be set to a common reference clock to be compared mathematically. We observed that ignoring these relativistic corrections introduces significant biases in the sky map results, particularly on the dipole. Ensuring a common time frame for $X$, $Y$, and $Z$—specifically, the \gls{tcb} associated with the \gls{bcrs} has alleviated this systematic bias and ultimately revealed the sought kinematic dipole, as will be detailed in the next section.
\begin{figure}
    \centering
    \includegraphics[scale=0.86, trim={0.61cm 0.7cm 0.0cm 0.0cm}, clip]{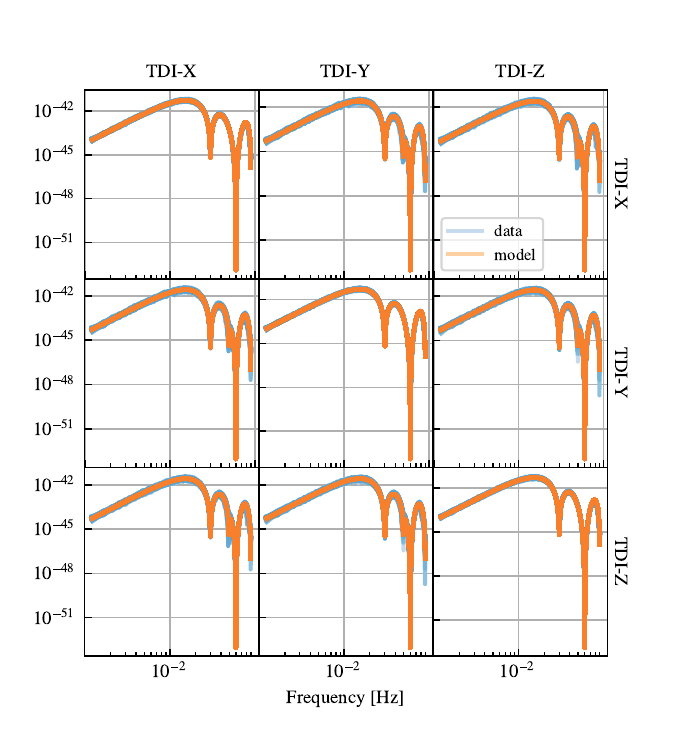}
    \caption{Averaged spectra of the $4$-years-long data set compared to the space interferometer response model to an anisotropic \gls{gw} sky. There are $N_t=384$ spectra over plotted for each in the figure, in {\it shades of blue} for the data, and in {\it shades of red} for the model.}
    \label{fig: data_vs_model_spectrum}
\end{figure}

\section{Data Analysis and map-making}
\label{section: data_analysis}

\subsection{The map-making likelihood function}
\label{subsec:map}
For our analysis, we employ 4 years of \gls{tdi} $2.0$ data streams sampled at $0.2 \,\si{\hertz}$. Initially, the data undergoes a pre-processing scheme involving frequency and time compression. The 4-year TDI data stream is partitioned into $N_t$ time steps, effectively determining the angular resolution of the analysis. Further data compression is achieved by averaging over frequency windows of width $n_j$. Based on the $X,Y,Z$ TDI channel data vector $\vb{d}$, one can determine the averaged data matrix as in \cite{baghi_uncovering_2023}
\begin{align}\label{equ:data_matrix}
    \mathbf{\bar{D}}(\tau_{i}, f_{j}) \equiv \frac{1}{n_{j}} \sum_{k=j-\frac{n_j}{2}}^{j+\frac{n_j}{2}} \vb{\tilde{d}}(\tau_{i}, f_k) \otimes \vb{\tilde{d}}(\tau_{i}, f_k)^{\dag}.
\end{align}
where the $N_t \times N_f \times 3$ data vectors $\vb{\tilde{d}}$ are the Fourier transforms of the time-split, simulated $X$, $Y$ and $Z$ time series. The matrix $\mathbf{\bar{D}}(\tau_{i}, f_{j})$ is the tensor product across \gls{tdi} channels of the data vectors $\vb{\tilde{d}}$, before averaging over the spectral window of width $n_j$ for data compression. It measures the cross-spectral density of the \gls{tdi} data streams. The statistical expectation of $\mathbf{\bar{D}}$ is the theoretical covariance $\vb C_d:= \vb C_{[X,Y,Z]}$ of the \gls{tdi} time series, which the Bayesian map-making method we use is ultimately solving for. The theoretical covariance $\vb C_d$ matrix can be computed, for each time and frequency bin, as
\begin{align}
    \vb{C}_{d}(\tau_{i},f_j) = \vb{A}_{d}(\tau_{i},f_j,p)\ I(f,p)\ + \vb{N}_{d}(\tau_{i},f_j),
\label{eq: covariance}
\end{align}
where $\vb{A}$ is the $N_t \times N_f \times 3 \times 3 \times N_{\text{pix}}$ matrix encoding the quadratic response function of the instrument \cite{lisa_cosmology_working_group_maximum_2020}, $I(f, p)$ is the \gls{gw} intensity sky map $N_{\text{pix}}$-vector, both showing explicit frequency and angular (pixel) dependence. Note that, as before, the product $\vb A_d I$ is summed over pixels. $\vb{N}_{d}$ is the $N_t \times N_f \times 3 \times 3$ covariance matrix of \gls{tdi} noise, and is optionally enabled in the analysis.
The intensity pixel map $I(p)$ contains the free parameter we aim to solve for in the analysis.\\
Aiming for a best-fit of the intensity map with respect to the input data, the appropriate probabilistic measure must be chosen based on the structure and statistical properties of the data vector $\vb{\tilde{d}}$. In this analysis, we employ a likelihood analysis based on the matrices $\mathbf{\bar{D}}(\tau_{i}, f_{j})$ statistics. In Equation (\myhyperref{equ:data_matrix}) we have seen that for each time segment $\tau$, $\mathbf{\bar{D}}$ is a $(3 \times 3)$ random matrix built from the average of the outer product of $n_j$ Gaussian distributed frequency series within the frequency window $\left[ j-\frac{n_j}{2}, j+\frac{n_j}{2} \right]$. It implies that $\mathbf{\bar{D}} \propto \vb{\tilde{d}} \vb{\tilde{d}}^\dag$ follows a Wishart distribution $\mathbf{\bar{D}}\sim W_{q=3}(n_j, \vb{C})$ \cite{SINHARAY201098}, with $\vb{C}$ is the covariance matrix of the Gaussian process $\vb{\tilde{d}}$. For $n_j \geq 3$ the probability density function\footnote{The density function is valid with respect to Lebesque measure on the cone of symmetric positive definite matrices.} of $\mathbf{\bar{D}}$ reads \cite{SINHARAY201098}
\begin{widetext}
\begin{align}\label{equ: wishart_density}
    f(\mathbf{\bar{D}}) = \frac{1}{2^{n_j q/2} \Gamma_q(\frac{n_j}{2})|\vb{C}|^{n_j/2}}|\mathbf{\bar{D}}|^{(n_j-q-1)/2}\exp{(-\frac{1}{2}\text{Tr}(\vb{C}^{-1}\mathbf{\bar{D}}))}.
\end{align}
\end{widetext}
Here, $\Gamma_q(\alpha)$ is the multivariate gamma function, and the dimension parameter $q = 3$ accounting for the three \gls{tdi} channels. Based on the latter equation, we can now formulate the likelihood $\mathcal{L}$ to maximize. Note that any constant prefactor will drop out in the logarithmic likelihood ratio computed by the MCMC sampling; hence, only the fitting parameters dependent factors are relevant.\\
For the purposes of this investigation, we define the log-likelihood function corresponding to a data sample $\mathbf{\bar{D}}$ as in Equation \eqref{equ:data_matrix}, i.e.
\begin{align}
    \log \mathcal{L} = \sum_{\tau_{i}}{\sum_{f_j}{ \bigg[ -\operatorname {tr} (\vb{C}_{d}^{-1}\mathbf {\bar D}(\tau_{i},f_j) ) - \nu \log |\vb{C}_{d}(\tau_{i},f_j)| \bigg] }},
\label{eq: logL}
\end{align}
where the trace is taken over the \gls{tdi} channels. We have introduced the effective number of \gls{dof} $\nu = \tfrac{n_j}{N_{bw}}$, accounting for the reduction factor $N_{bw}$, also known as the normalized equivalent noise bandwidth \cite{heinzel_spectrum_2002}, which depends on the time segments $\tau$ overlap and the window functions applied when time splitting the data \footnote{we refer the reader to the comprehensive review \cite{heinzel_spectrum_2002} for further details}. \\
It is evident that the fitting parameters enter via the covariance matrix $\vb{C}_d$ and, within it, via the intensity map. The Bayesian map-making strategy necessitates the numerical computation of the covariant matrix $\vb{C}_d$ for each step in the \gls{mcmc} parameter space exploration, employing the left-hand side of Equation \eqref{equ:pix_vs_smooth}. While the evaluation of the $\log \mathcal{L}$ function requires the intensity pixel map $I_p$, several methods can be employed for extracting the relevant information from $I_p$ using a minimal set of fitting parameters. For our investigations, we focus on the following two:\\
The first choice relies on the assumption that the anisotropies in the signal are produced by a Doppler boost. For a scale-invariant energy density spectrum, as given in Equation \eqref{equ:1}, a suitable reference frame can be fixed without loss of generality, leaving four parameters that determine the pixel map $I_p$, i.e. the monopole power in the observer frame, $\Omega_{GW}'(f)$, and three velocity components $\va{\beta} = [\beta_x, \beta_y, \beta_z]$, as appearing in \eqref{equ:1}. These components are related to the boost vector of the solar system relative to the \gls{cmb} frame and are expressed in the \gls{bcrs}-frame. Combining Equations \eqref{equ:0} and \eqref{equ:energydensity} yields the intensity $I_p$ and, correspondingly, $\vb{C}_d$. The advantage of this model lies in its small number of fit parameters, significantly reducing the numerical complexity of our pipeline. However, this technique is fully model-dependent, meaning it is only valid for Doppler-sourced anisotropies originating from a scale-free source being the dominant contribution in the cosmological \gls{sgwb}.\\
Another approach of extracting the intensity map from a given model is through its decomposition into modes of the spherical harmonics' basis. In this scenario, the pixel map $I_p$ is decomposed into 6 independent modes (considering only up to quadrupolar anisotropies, i.e. to $\ell = 2$), where the $\ell, \pm m$ modes are interdependent since $I_p$ is real for every pixel. The 6 modes correspond to 9 independent parameters, as 3 of the relevant modes are complex. This decomposition is model-independent and, in principle, captures any anisotropy within the data, including intrinsic anisotropies in the source frame.\\
Note that both parametrization discussed above are only valid for a scale free spectrum. For more complex spectra, i.e. spectra with non-trivial spectral dependency, knowledge of the corresponding fit parameters is required for each frequency bin individually.

\subsection{MCMC sampling and map reconstruction}
\label{subsec:mcmc}

In our investigation of the two parameter spaces, we seek the optimal fit for the intensity map, denoted as $I_p$, using simulated \gls{lisa} data through \gls{mcmc} sampling of the likelihood function outlined in Equation \eqref{eq: logL}. Despite the computational demands associated with exploring the parameter space via \gls{mcmc}, this approach offers a notable advantage by circumventing the numerically challenging Fisher matrix inversion. The latter inversion, when employed, can introduce systematics and compromise the overall robustness of the analysis \cite{lisa_cosmology_working_group_maximum_2020, bartolo_probing_2022}. Moreover, beyond the numerical challenges, the analytical derivation of the Fisher iteration typically involves a Gaussian approximation, assuming the likelihood function approximates a Gaussian distribution near its peak \cite{bond_estimating_1998}.
In contrast, our \gls{mcmc} scheme is not relying on any analytical or numerical approximations of the log-likelihood function in the Equation \myhyperref{eq: logL}. As our focus is confined to low modes in the study of kinematic anisotropy, an \gls{mcmc} map-making strategy based on the methodology presented in \cite{lisa_cosmology_working_group_maximum_2020} becomes particularly pertinent. Furthermore, the inherent low angular resolution of \gls{lisa}, limiting the resolution of higher modes, renders an \gls{mcmc} map-making approach a practical and effective solution for many scenarios encountered in \gls{lisa} analyses.\\
Our mapping scheme is structured as follows: At each iteration of the algorithm, the likelihood of Equation \eqref{equ:0} is evaluated based on the current state's location in the parameter space. In principle, the covariance matrix can be calculated by inserting Equation \eqref{equ:1} into \eqref{eq: logL}; however, numerically we are not bound to the necessity of expanding in small velocities and can instead use the full expression, \eqref{equ:0}, to calculate the $\vb{C}_d$ and the likelihood correspondingly. As introduced in section \ref{sec:doppler}, for a flat and isotropic energy density spectrum $\Omega_{GW}(f)$ in the \gls{cmb}-frame, the Doppler shift due to the observer's velocity with respect to stochastic emission induces a sky modulation of its amplitude that does not depend on the frequency, i.e. $\alpha_\Omega= n_\Omega = 0$. This implies that changing the velocity vector $\va{\beta}$ will have an overall effect on the sky map at all frequencies, and the treatment of frequency and angular dependence remains separate,
\begin{align}\label{equ: sim cov}
    \vb{C}(\tau,f) & = \vb{A}(\tau,f,p) \ I(f,p) \nonumber \\
                & = \vb{A}(\tau,f,p) \ \frac{E(f)}{E(f_0)} \ I(f_{0}, p)
\end{align}
where we have extracted  the spectral dependency of the energy density in the \gls{cmb}-frame via the function $E(f)$ as in \cite{lisa_cosmology_working_group_maximum_2020}, and we use the (arbitrary) reference frequency of $f_0 = 1\ \si{\milli\hertz}$ to define the intensity map $I(p) = I(f_{0}, p)$ we are fitting for. Thus, for each iteration $i$, the evaluation of Equation \eqref{equ:0} boils down to computing the prefactor \eqref{equ:-1} for the new velocity vector $\vec \beta^i$ when applying the model-dependent fit ansatz. Given the fit parameters $\{\Omega^i_{GW},\beta^i_x, \beta^i_y, \beta^i_z\}$ of iteration $i$ selected by an arbitrary walker of the \gls{mcmc} the algorithm proceeds by calculating 
\begin{align}
     I^i(f_0,p) &= \frac{3 H_0^2}{32 \pi^3 f_0^3} \mathcal{D}^{4}(p) \ \Omega_{GW}^i\left( f_0 \right) \label{equ: HI and the BW ghost}
\end{align}
with
\begin{align}
     \mathcal{D}(p) &= \frac{\sqrt{1-|\vec\beta^i|^2}}{1 - |\vec\beta^i| \  \left[ \vu{k}\cdot\vu{\beta}^i \right]_{p}}. 
\label{equ: sim omega}
\end{align}
The angular or pixel dependence of $I(f_0,p)$ is introduced implicitly through the Doppler-boosting of $\Omega_{GW}^i$ from the source into the observer frame via $\mathcal{D}$. Note that in the case of a slope-free spectrum, such as for \gls{cs}, $\Omega_{GW}(f) = \Omega_{GW}$ is constant. \\ The resulting discrete $I^i(f_0, p)$ is then summed over pixels, as in Equation \eqref{equ: sim cov}, yielding the covariance matrix $\vb{C}^i$ for the $i$-th iteration. With the newly acquired covariance matrix, the likelihood function \eqref{eq: logL} is computed, and the results are compared to those of the other walkers within the corresponding iteration. By the definition of the scheme, the walkers statistically favor the vicinity around high-likelihood points in parameter space. Eventually, they converge toward a maximum of \eqref{eq: logL} for the given input data. In principle, the \gls{mcmc} can be initialized in an arbitrary state. Including any previous knowledge about the anticipated signal will most likely contribute to the scheme converging towards the correct maximum likelihood faster. Here, we favor an agnostic attempt, initializing the \gls{mcmc} in an isotropic map configuration $\Tilde{I}_p = \text{constant}$. For a detailed description of the specific \gls{mcmc} scheme employed in this work, we refer to \cite{foreman-mackey_emcee_2013, goodman_ensemble_2010}.
\begin{figure}
    \centering
    \includegraphics[width=\linewidth, trim={2.2cm 2.3cm 0.0cm 0.5cm}, clip]{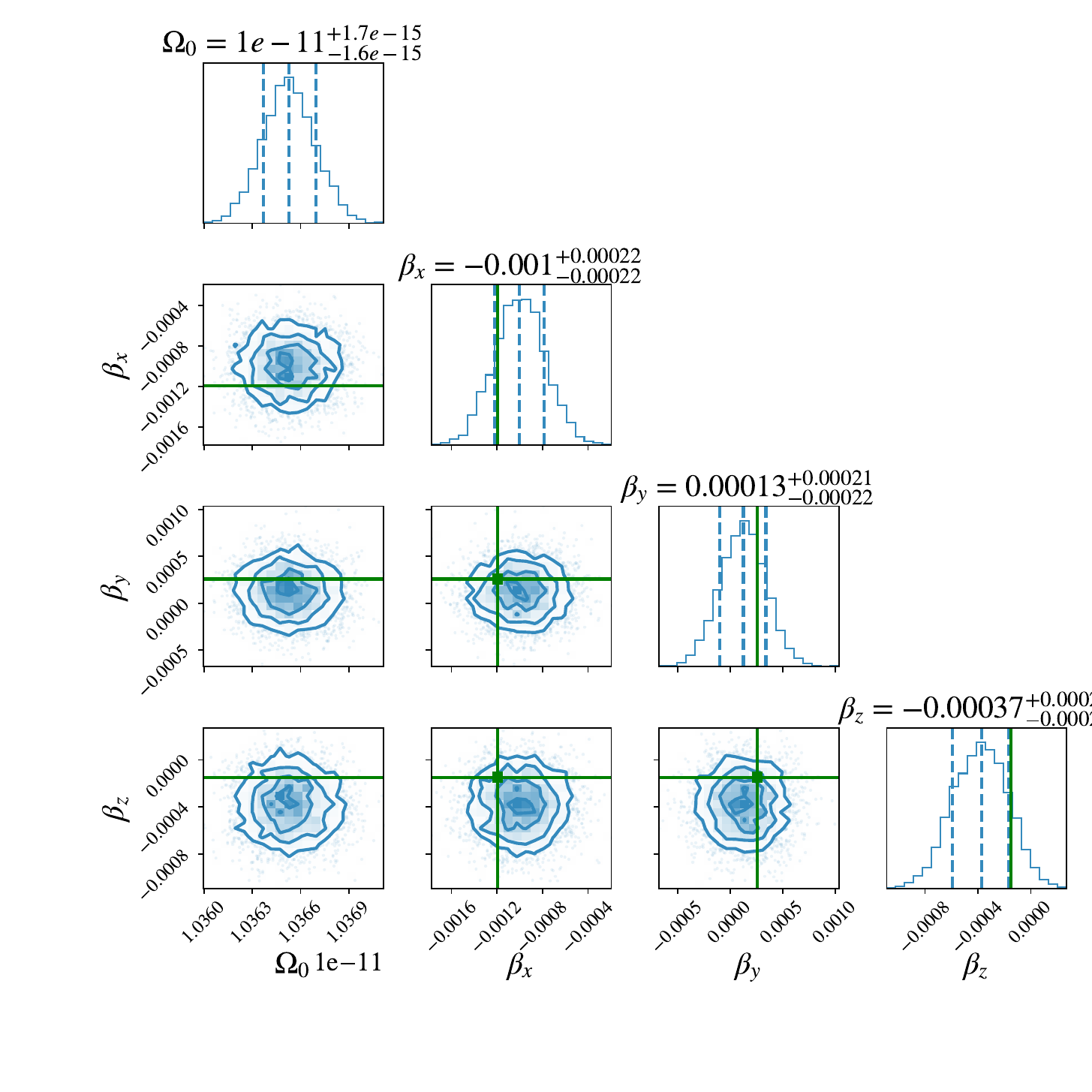}
    \caption{Samples histogram of the $(\Omega_{GW},\Vec{\beta})$ space for a single $4$-years-long \gls{gw} sky realization after convergence of the \gls{mcmc}.}
    \label{fig: cornerplot}
\end{figure}
\begin{figure*}
    \centering
    \includegraphics[width=\textwidth, trim={0.0cm 40.0cm 0.0cm 0.0cm}, clip]{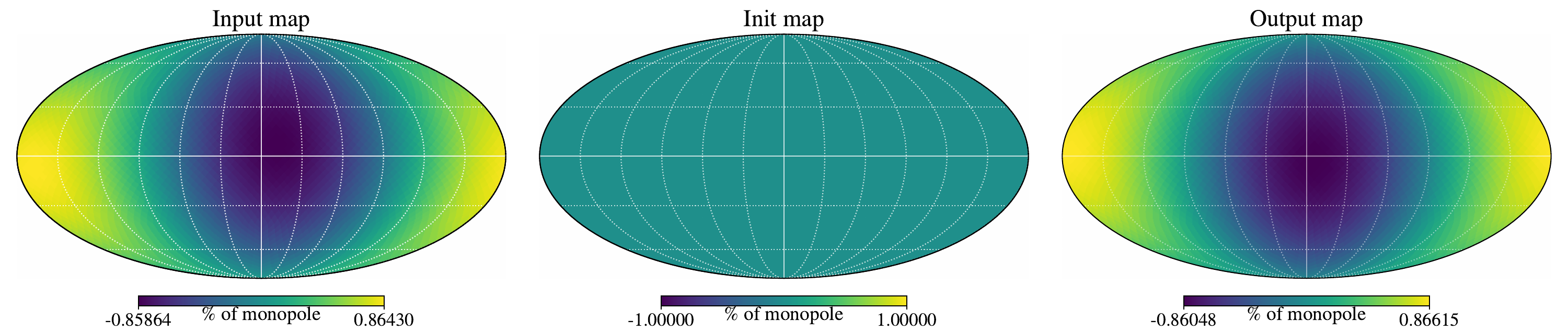}
    \caption{Intensity \gls{gw} sky maps (in percentage of the monopole $\Omega_0$). From left to right: the input intensity map injected, the a priori map used to initialize the \gls{mcmc} chains, and the map recovered by the map-making algorithm. The monopole is removed from the intensity maps, to ease the visualization the dipolar anisotropy. These projection plots have been made from \texttt{healpy} python package.}
    \label{fig: skymaps}
\end{figure*}\\
In Figure \ref{fig: cornerplot}, we display an exemplary histogram and correlation diagram of the \gls{mcmc} chains after convergence, focusing on a single realization of the \gls{sgwb} sky. The \gls{mcmc} chains exhibit convergence towards the true velocity $\beta^{\text{true}}$. The statistical properties of the \gls{mcmc} samples, capturing the posterior distribution, offer a theoretical estimate of the measurement precisions. For this specific sky realization, $\beta^{\text{true}}$ is observed to be more than $4$-$\sigma$ away from zero, indicating the detection of the observer's motion with respect to the \gls{cmb} rest frame. Figure \ref{fig: skymaps} reconstructs the recovered skymap from the $\beta^{\text{meas}}$ measurement in Figure \ref{fig: cornerplot}. We verified an excellent match between the input and output intensity maps, considering that the prior, starting point sky map was perfectly isotropic.\\
While the results from Figures \ref{fig: cornerplot} and \ref{fig: skymaps} suggest the feasibility of resolving kinematic anisotropies with the \gls{mcmc} map-making strategy, it is essential to acknowledge that, at this stage, the analysis remains susceptible to potential statistical fluctuations in the input sky map realization, leading to a fortuitous enhancement of the dipolar signal far from its expectation value. We refer to this signal-induced statistical fluctuation of the $\ell=1,2$ kinematic modes as ``cosmic variance'', drawing an analogy to the term as defined in the context of the \gls{cmb}, where the measured positions of the peaks in the spectrum are limited by the fact that there is only a single realization of the spectrum observable from Earth \cite{White_1993}. Additionally, it is crucial to consider the possibility of an unfortunate systematic in the analysis that could have contributed to the observed signal.

\section{Analysis results}
\label{sec:analysis}
In the following section, we will delve further into investigating the analysis artifacts discussed above and demonstrate that they can be excluded as plausible sources contributing to the observed signals in Figures \ref{fig: cornerplot} and \ref{fig: skymaps}.\\
To test against cosmic variance, we generate 30 independent realizations of the \gls{gw} sky and reiterate the map-making procedure outlined in Section \myhyperref{subsec:mcmc} for each realization in \myhyperref{subsection: cosmic variance}. Subsequently, in section \myhyperref{subsection: systematic}, the input sky map is rotated arbitrarily to verify the congruence between the injected and recovered intensity maps, excluding systematic errors as an explanation for the observed signal. To achieve this analysis, we introduce instrumental noise to the simulation to assess \gls{snr} and discuss the detectability of scale-free \gls{sgwb} kinematic anisotropies for \gls{lisa} in section \myhyperref{subsection: noise}.

\subsection{Statistical significance and cosmic variance}
\label{subsection: cosmic variance}

\begin{figure}[h!]
    \centering
    \includegraphics[width=\columnwidth, trim={0.55cm 0.3cm 0.05cm 0.0cm}, clip]{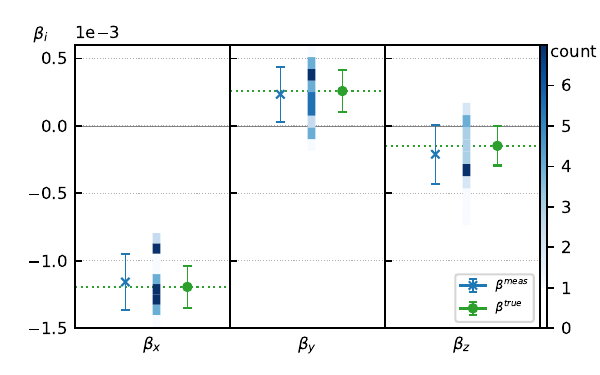}
    \caption{Variance of the measured signal over the $30$ realizations of the \gls{gw} sky (blue histograms), compared to the resulting averaged values and standard deviations (in blue), and  the theoretical values and \gls{mcmc} error bars (in green). Are plotted the data for $\beta_x$, $\beta_y$ and $\beta_z$, from left to right. The dispersion of $\beta^{\text{meas}}$ realizations are statistically consistent with the true values $\beta^{\text{true}}$ and the theoretical error bars from the \gls{mcmc} sampling.  $\beta_x$ is resolved $> 4 \sigma$ away from $0.0$ value.}
    \label{fig: stats_beta}
\end{figure}

\begin{figure}
    \centering
    \includegraphics[width=\columnwidth, trim={0.8cm 0.5cm 2.0cm 0.0cm}, clip]{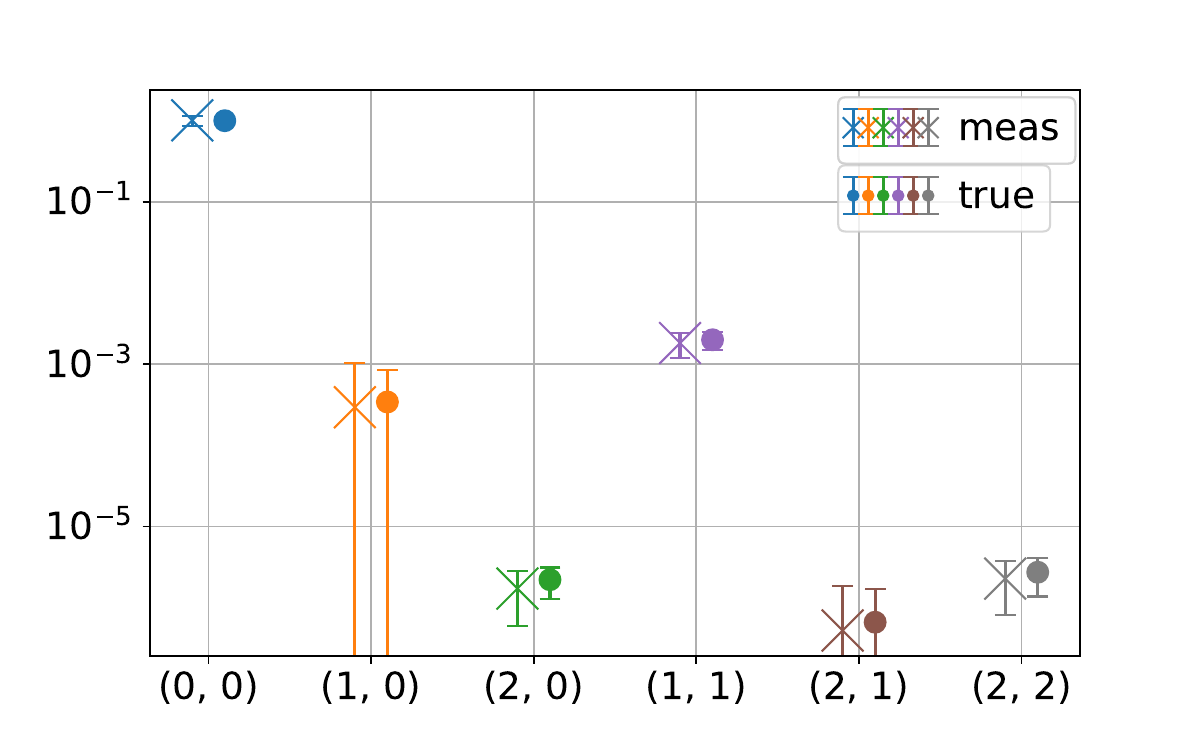}
    \caption{Variance of the measured signal against the theoretical error bars for the mode components of the \gls{sgwb} in the observer frame up to $\ell=2$ based on the statistics for $\beta_i$, Figure \ref{fig: stats_beta}.}
    \label{fig: stats_alms}
\end{figure}

The statistical robustness of the signal depicted in Figures \myhyperref{fig: cornerplot} and \myhyperref{fig: skymaps} is assessed through a comparison of $30$ randomly generated realizations of a stochastic \gls{gw} sky. Figure \ref{fig: stats_beta} illustrates the dispersion of the measured velocity $\beta^{\text{meas}}$ in the \gls{sgwb} rest frame with respect to each distinct realization of the \gls{sgwb}. The figure also presents a comparison between the mean value and standard deviation over the $30$ realizations, alongside the true velocity values and theoretical uncertainty estimates from the \gls{mcmc} sampling. The analysis in Figure \ref{fig: stats_beta} shows that the statistics of the $30$ measurements align well with the theoretical values. Specifically, the average $\overline{\beta}^{\text{meas}}$ converges towards $\beta^{\text{true}}$, and the measured standard deviation is compatible with the theoretical error bars. Notably, the average measured $\overline{\beta}_x^{\text{meas}}$ is well resolved and located $> 4 \sigma$ away from $0$. This numerical demonstration establishes that even in the challenging and poor-featured scenario of a boosted, scale-free \gls{cs}-like stochastic signal, the inherent {\it cosmic variance} is effectively overcome.
Based on the $30$ realizations, we also compute the associated intensity maps and their $a_{\ell m}$ components in the spherical harmonics decomposition. In Figure \myhyperref{fig: stats_alms}, we conduct the same statistical studies to verify the robustness of the resolution of kinematic spherical harmonics mode and observe a clear measurement of the dipolar, unambiguous $(1,1)$ mode, which constitutes the principal component of the boosted \gls{sgwb} data with a flat energy density spectrum. Here, the $\ell=1$ modes are, as anticipated, orders of magnitude brighter than the $\ell=2$ modes, given their intensity scales as $\beta^{\ell}$. However, notably, the \gls{mcmc} starts to exhibit sensitivity to quadrupolar components such as the $(2,0)$ and $(2,2)$ modes, detected here approximately with an $ 2 \sigma$ confidence level. This sensitivity arises from the fact that \gls{lisa} is significantly more responsive to quadrupolar signals than to dipolar signals due to the pronounced parity of its response \cite{Doppler_effort_IV} \cite{bartolo_probing_2022}. This enhanced response partially compensates for the intrinsic weakness of the quadripolar component, bringing it within the observational reach of \gls{lisa}. In some cases, cosmological signals with richer spectral signatures can further enhance the $\ell=2$ component, and it is possible that $\ell =2$ modes may become comparable or even the dominant kinematic signatures in the observed signal.

\subsection{Systematic check}
\label{subsection: systematic}
Having established the statistical significance of the result in Section \myhyperref{subsec:mcmc}, we now address the possibility that the observed signal may be attributed to biases or systematic errors in the instrument response or the analysis. To scrutinize this aspect, we introduce a Doppler-induced anisotropic sky map by injecting a rotated velocity vector $\va{\beta}^{\text{rot}}$. Subsequently, we verify our ability to recover this modified injection. The new velocity vector to be recovered is given by $\va{\beta}^{\text{rot}} = R_{ZYX}(\phi, \eta, \theta)\ \va{\beta}^{\text{true}}$, where the rotation matrix $R_{X,Y,Z}$ contains the angles $\phi = 100 \si{\degree}$, $\eta = 0 \si{\degree}$, $\theta = 60 \si{\degree}$ representing the Euler rotation angles around the $Z$, $Y$ and $X$ axes (of the \acrshort{ssb} frame) respectively.
\begin{figure}
    \centering
    \includegraphics[width=\columnwidth, trim={0.45cm 0.3cm 0.85cm 0.0cm}, clip]{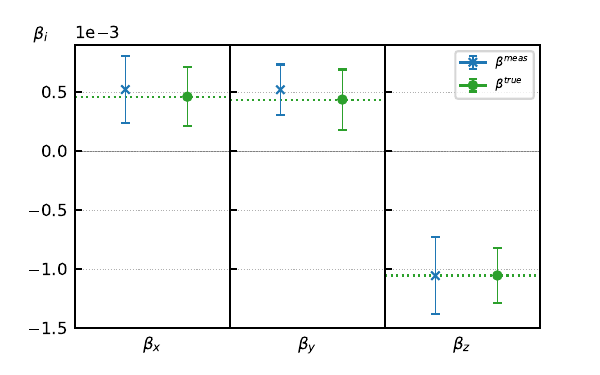}
    \caption{Variance of the measured signal over the $10$ realizations of the {\it rotated} \gls{gw} sky (blue histograms), compared to the resulting averaged values and standard deviations (in blue), and  the theoretical values and \gls{mcmc} error bars (in green). Are plotted the data for $\beta_x^{\text{rot}}$, $\beta_y^{\text{rot}}$ and $\beta_z^{\text{rot}}$, from left to right. We do verify that new, modified input velocity $\beta^{\text{rot}}$ is recovered with identical precision that in Figure \myhyperref{fig: stats_beta}}
    \label{fig: stats_beta_rotated}
\end{figure}
Figure \myhyperref{fig: stats_beta_rotated} displays the injected values $\va{\beta}^{\text{rot}}$ and the recovered values for $10$ realizations of the sky. Similar to Figure \myhyperref{fig: stats_beta}, the analysis successfully captures this newly injected signal. This observation demonstrates that the presented analysis pipeline can accurately reconstruct the injected sky map, providing evidence that the measured intensity map in Figure \myhyperref{fig: skymaps} is not attributable to a systematic error.

\subsection{Model-independent analysis}
\label{subsection: model-independent}
\begin{figure}
    \centering
    \includegraphics[width=\columnwidth, trim={0.8cm 0.5cm 2.0cm 0.0cm}, clip]{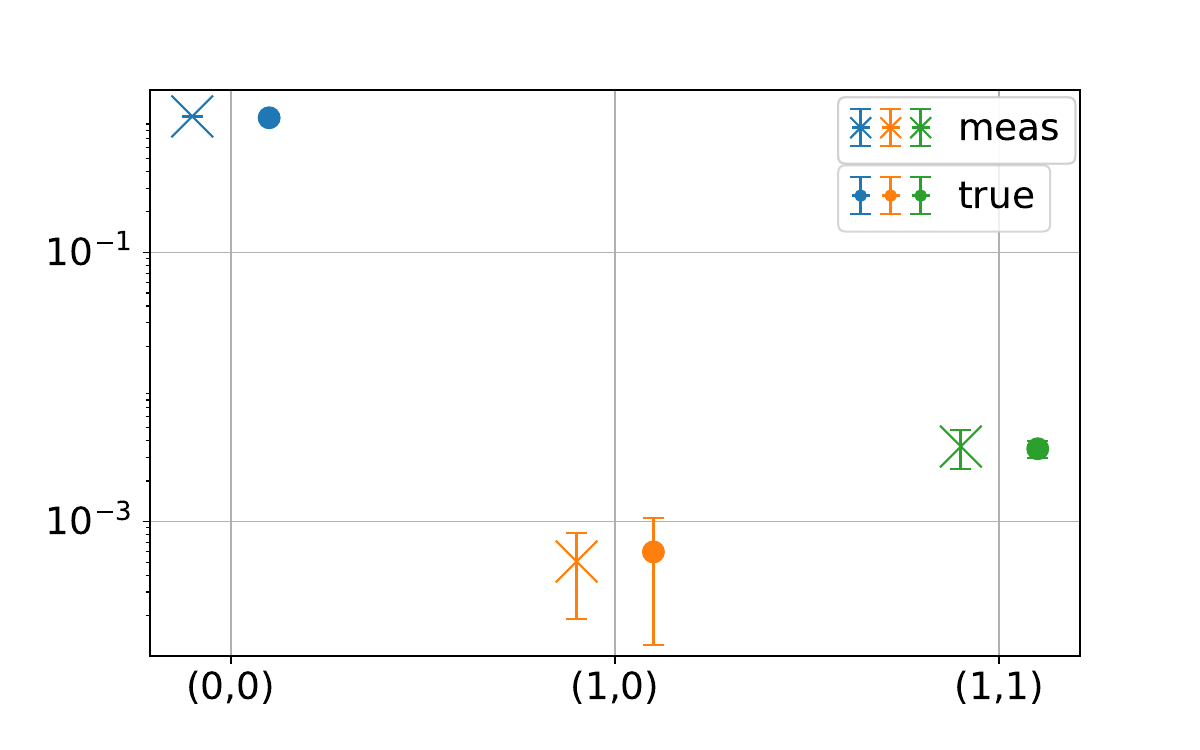}
    \caption{Variance of the measured signal against the theoretical error bars for the mode components of the \gls{sgwb} in the observer frame up to $\ell=1$ based on a \gls{mcmc}-sampling using $a_{\ell m}$.}
    \label{fig: stats_alms_alms}
\end{figure}

So far, the analysis focuses on the model-dependent ansatz for the choice of the fit parameters, i.e. we were fitting with respect to $\{\Omega_{GW}, \vec \beta\}$. We now switch to the fully agnostic approach and employ a fit for the spherical modes of the signal, parameterizing the intensity map $I_p$ using its mode decomposition $i_{\ell m}$, see \eqref{equ:mode_decomp}, up to $\ell\leq 2$. Note that under the assumption that the signal is of purely kinematic nature, i.e. results from the Doppler-boost of an isotropic background, the set $\{\Omega_{GW}, \vec \beta\}$ represents the minimal number of parameters to characterize the signal. In a general setting, however, a model-independent analysis enhances the versatility of the outlined pipeline. In contrast to the model-dependent ansatz, using the modes $i_{\ell m}$ enables the resolution of any angular dependence in the signal $I_p$, including both intrinsic or Doppler-induced sky anisotropies. \\
As before, we analyze the model-independent choice of fit parameters for $10$ random implementation of a \gls{cs}-like \gls{sgwb} signals. For each iteration $k$, the intensity $I^k_p$ is computed via the modes $i_{\ell m}^k$ selected by the \gls{mcmc} based on the likelihood function \eqref{eq: logL}. The results are displayed in Figure \ref{fig: stats_alms_alms}. We observe that for $\ell=0,1$, the model-independent approach matches the accuracy of the parametrization $\{\Omega_{GW}, \vec \beta\}$. The measured mean values $\bar i_{\ell m}$ are statistically significant and exhibit a variance comparable to the theoretical error bars. Allowing for $\ell=2$, the \gls{mcmc} converges towards an undesirable maximum, largely overestimating the monopolar and quadrupolar contributions. This is partially due to the correlation between monopole and quadrupole manifesting in an equal power in $\beta$ carried by both, see Equation \eqref{equ:M} and \eqref{equ:Q}, and which is is confirmed numerically. We are lead to the conclusion that, due to the correlation of $M$ and $Q$, the model-independent approach allows for statistical uncertainties of the monopole to leak into the quadrupole. The consequence of this leakage resides in mono- and quadrupole reaching similar orders of magnitude in the output skymap $I_p^{\text{out}}$. This undesirable effect can be counteracted by fine-tuning analysis parameter such as frequency- and time-binning, or disentangling mono- and quadrupole via different parametrizations. One could also think of recursive schemes by which we first fix $\ell=0,1$ via the \gls{mcmc} before fitting for $\ell=2$ only. All options have been considered, however no optimal strategy has been identified so far. We leave a deeper investigation to future works.

\subsection{Instrumental noise and detectability prospects of boosted scale-free \gls{sgwb}}
\label{subsection: instrumental noise}
\begin{figure}[]
    \centering
    \includegraphics[width=\columnwidth, trim={0.0cm 0.0cm 0.0cm 0.0cm}, clip]{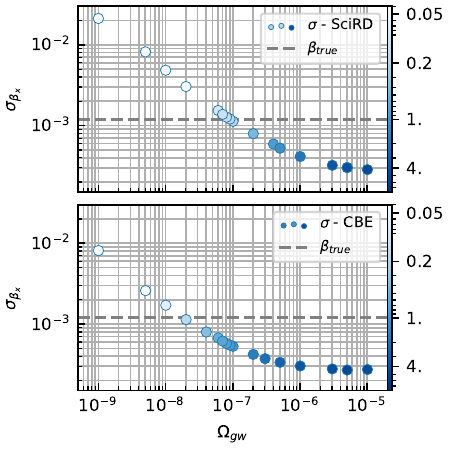}
    \caption{Analysis of the evolution of $\sigma_{\beta_{x}}$ as the signal amplitude $\Omega_{GW}$ varies. The blue color bar provides the estimated standard deviation $\tfrac{\sigma_{\beta_{x}}}{\mid \beta_{x}^{\text{true}} \mid}$ in $\mid \beta_{x}^{\text{true}} \mid$ unit. Two instrumental noise models are considered: \gls{scird} specifications (top) and current best estimate (down), differing essentially by the level of \gls{oms} noise (displacement noise floor at $1.5 \times 10^{-11}$ and $6.35 \times 10^{-12} \si{\metre\persqrthz}$ resp. \cite{Scird, bayle_unified_2023}). Both cases confirm that the kinematic dipole is reachable for $\Omega_{GW}$ values typically above $10^{-7}$. We observe that for $\Omega_{GW} > 10^{-7}$ we converge to the cosmic variance limit displayed in Figure \myhyperref{fig: stats_beta}.}
    \label{fig: noise_analysis}
\end{figure}

At this stage, instrumental noise has not been incorporated. The inherent stochastic characteristics of the signal, specifically the intrinsic variance of its spectrum density competing with its annual temporal fluctuations, has constituted the exclusive sources of randomness and uncertainty under consideration. As illustrated in Figure \myhyperref{fig: stats_beta}, the precision of the map-making process surpasses the anticipated cosmic variance for a data span of $4$ years. Importantly, this outcome has remained entirely unaffected by the signal amplitude $\Omega_{GW}$ in the noise-free case, which has not been a subject of discussion hitherto.\\
To investigate the detectability (i.e. including instrumental noise) of Doppler boost-induced anisotropies in a scale-free extra-galactic \gls{sgwb}, we now take the instrumental noise into consideration. We adopt a zero-noise likelihood \gls{mcmc} map-making approach. In this methodology, we calculate the covariance matrix (\myhyperref{eq: covariance}) by considering an expected value for the noise matrix $\vb{N}_d$ derived from the noise model detailed in \cite{baghi_uncovering_2023}. While the data matrix $\vb{D}$ in the log-likelihood definition (\myhyperref{eq: logL}) remains unaltered, the modeled covariance matrix $\vb{C}_d$ now incorporates instrumental noise, influencing the uncertainties associated with $\va{\beta}^{\text{meas}}$ assessed through the \gls{mcmc} analysis. Figure \myhyperref{fig: noise_analysis} presents the results of this analysis. In contrast to the approach in \cite{baghi_uncovering_2023}, we consider two noise performance scenarios: one conservatively aligned with the LISA \gls{scird} \cite{Scird}, and a second more optimistic scenario in accordance with our current best understanding of the instrument's performance (see Appendix B and Eq. (B5) in \cite{bayle_unified_2023}). The figure illustrates the estimated $\sigma^{mcmc}_{\beta_x}$ as a function of $\Omega_{GW}$, with the color scale indicating, for each data point, the distance to the target $\beta_x^{\text{true}}$ in units of $\sigma^{mcmc}_{\beta_x}$.
\label{subsection: noise}
\\
Figure \myhyperref{fig: noise_analysis} affirms the findings reported in \cite{bartolo_probing_2022}. In the top subplot, under a conservative noise configuration, we find that the observability of the kinematic dipole becomes significant only at relatively large values of $\Omega_{GW} \gtrsim 10^{-7}$. The more optimistic scenario, utilizing the latest LISA performance model, indicates detectability at $\Omega_{GW} \gtrsim 5 \times 10^{-8}$, potentially reaching $10^{-8}$. This order of magnitude aligns with expectations, considering LISA's suboptimal sensitivity to odd-$\ell$ modes \cite{bartolo_probing_2022}. \\
With this comprehensive analytical tool at our disposal, we are poised to explore avenues for improving detectability. This involves optimizing the preprocessing of the TDI data streams through time-frequency analysis, considering more realistic orbital parameters, or leveraging spectral features in the \gls{sgwb} to enhance both dipolar and quadrupolar kinematic anisotropies.

\section{Summary and Discussion}
\label{sec:discussion}
In this article, a full time-domain map-making scheme based on a maximum-likelihood \gls{mcmc} tailored to \gls{lisa} and targeting kinematic anisotropies in the \gls{sgwb} is presented. We generated stochastic strain time series for each pixel of the sky and projected it to the simulated, full instrument response function. For the resulting data streams consisting of channeled random data matrices convoluted with the instrumental response kernel, we selected a suitable model-based likelihood function to test the data against a physical model. Based on the selected likelihood function, a novel \gls{mcmc} mapping scheme was developed. The efficacy of our pipeline was demonstrated by successfully recovering the injected map of Doppler-boosted primordial anisotropic data, showcasing
the functionality of the \gls{mcmc}-based mapping scheme
for \gls{lisa}. Our results indicate that our pipeline surpasses
cosmic variance limitations in \gls{lisa} data and exhibits robustness against systematical and statistical errors. The recovery of injected data was achieved using both model-dependent and independent sets of fit-parameters. Incorporating instrumental noise, we determined that the pipeline converges towards a statistically meaningful recovery of the kinematic dipole for $\Omega_{GW}\gtrsim 10^{-7}$ when conservative noise scenarios are applied, and slightly better for more favorable instrumental noise realizations\footnote{Here, $\Omega_{GW}$ refers to the monopole power of the isotopic signal in the source frame.}. These findings align with and complement the results reported in \cite{bartolo_probing_2022}.\\
\\
Despite these promising results of our analysis, we acknowledge areas for improvement that will be incorporated in the pipeline in future works. Specifically, we identified opportunities to optimize the data sampling process by switching to a time-frequency domain instead of a full, time-domain analysis. Replacing the short-Fourier transform with a wavelet-based analysis in the preprocessing phase can enhance the likelihood of recovering kinematic anisotropy from the data. Furthermore, we observed that the orbit dynamics have a significant impact on the output of the map-making scheme, suggesting that simulating more realistic orbit dynamics, as opposed to the static equal-arm-length assumption applied here, helps to enrich the kinematic signatures in the data. We have observed, however, that the use of more realistic keplerian orbits introduce important biases in the map-making analysis, which are still under investigation. They seem to be tightly related to the initial preprocessing step, where the $4$ years \gls{tdi} data streams are time split using $50\%$ overlapping Haning windows. We believe that moving to a time-frequency representation of the \gls{tdi} data will help fixing this issue, in the aim of exploiting the complex temporal features the full orbital dynamics provides to the response. More generally, we wish to address the sensitivity of the proposed map-making scheme to relevant parameters of the analysis, in particular the number of time- and frequency bins, $n_{\text{samples}}$ and $n_{\text{avg}}$, in which the input data is divided\footnote{Compare with Equation \eqref{equ:data_matrix} and subsequent passages.}. We found that the convergence of the \gls{mcmc} reacts sensitively to the choice of $n_{\text{samples}}$ and $n_{\text{avg}}$. In particular, the monopole can be offset significantly for suboptimal choices of bin size. For the results displayed in the previous section, the parameters $n_{\text{samples}}$ and $n_{\text{avg}}$ were optimized with respect to the random walkers' convergence in the parameter space. While the implications of a shifted time (or frequency) grid within the context of the \gls{mcmc} are not yet fully understood, they are currently being investigated and will be subject to future works. Again, we expect that the use of wavelet transforms will facilitate these investigations. \\
Finally, it is imperative to address more intricate primordial contributions to the \gls{sgwb} exhibiting a non-trivial spectral dependency, see e.g. Figure \ref{fig:spectrum}. In the analysis above, we only considered a scale free spectrum motivated by CS contributions. However, extending this analysis to richer signatures can significantly improve the lower bound on $\Omega_{GW}$ for which the \gls{mcmc} recovers the desired contribution. This statement is attributed to the dependence of the components $D(f)$ and $Q(f)$ on the slope, as outlined in Equations \eqref{equ:D} and \eqref{equ:Q}. Both functions, effectively representing dipole and quadrupole modes of the signal, exhibit notable amplifications when the slope of the spectrum $\Omega_{GW}$ varies with frequency. This variation is quantified by the functions $n_\Omega$ and $\alpha_\Omega$. In certain instances, this can result to the quadruple power surpassing that of the dipole. Consequently, spectra originating from phenomena such as, for instance, PTs or PBHs strongly amplify the power of the kinematic anisotropies for $\ell>0$, with a notable emphasis on enhancing the quadrupole. The augmentation of signal power for $\ell = 2$ is particularly significant for LISA, as the instrument exhibits heightened sensitivity to the quadrupolar mode of sky anisotropies \cite{bartolo_probing_2022}.\\
\\
Aside from the identification of cosmic origins of signatures in \gls{lisa} data, our pipeline can be exploited to analyses galactic contributions to the \gls{sgwb} as well. In this work, we build a framework assuming that an effective foreground removal scheme has been applied to the raw data. This step is highly non-trivial but falls out of the scope of this article. However, our map-making algorithm can very well be exploited for the mapping of the galactic confusion noise.
The capability of the developed \gls{mcmc}-mapping scheme to map the remaining Milky Way \gls{gw} sources in \gls{ldc} data after extracting resolved binaries will be addressed in a follow-up project. \\
In conclusion, this analysis presents a ready-to-use toolkit enabling tests of suitably processed \gls{lisa} data of \gls{sgwb}-nature for signatures of cosmic origin. Concretely, given the \gls{lisa} sensitivity band being swamped by a \gls{sgwb} signal, signatures rich in spectral features may be recovered for $\Omega_{GW}\gtrsim 10^{-8}$ given a less conservative instrumental noise realization. This utility becomes crucially relevant in the case of possible PTA signatures of a stochastic background \cite{agazie_nanograv_2023_II} extending into the LISA band. If the latter signal is (partially) of cosmic origin, one expects very relevant contributions of the same source in LISA data band. A cosmic character of these hypothetical signatures can be established using the here presented analysis pipeline. 

\section*{Acknowledgements}
LH would like to acknowledge financial support from the European Research Council (ERC) under the European Unions Horizon 2020 research and innovation programme grant agreement No 801781. LH further acknowledges support from the Deutsche Forschungsgemeinschaft (DFG, German Research Foundation) under Germany’s Excellence Strategy EXC 2181/1 - 390900948 (the Heidelberg STRUCTURES Excellence Cluster). The authors thank the Heidelberg STRUCTURES Excellence Cluster for financial support. The authors acknowledge support by the state of Baden-Württemberg, Germany, through bwHPC. Computations were also performed on the DANTE platform, APC, France. The authors would like to thank Q. Baghi, J-B. Bayle, M. Besançon and N. Karnesis for their precious insight and the numerous fruitful discussions we had along this project. The authors also thank the LISA Simulation Working Group and the LISA Simulation Expert Group for the lively discussions on all simulation-related activities.

\printglossary[type=\acronymtype]
\bibliographystyle{unsrt2}
\bibliography{add_ref}
\end{document}

%% file: acronym.tex
    
\newacronym{e2e}{E2E}{End-To-End}
\newacronym{inrep}{INREP}{Initial Noise REduction Pipeline}
\newacronym{tdi}{TDI}{Time Delay Interferometry}
\newacronym{ttl}{TTL}{Tilt-To-Length couplings}
\newacronym{dfacs}{DFACS}{Drag-Free and Attitude Control System}
\newacronym{ldc}{LDC}{LISA Data Challenge}
\newacronym{lisa}{LISA}{the Laser Interferometer Space Antenna}
\newacronym{emri}{EMRI}{Extreme Mass Ratio Inspiral}
\newacronym{ifo}{IFO}{Interferometry System}
\newacronym{grs}{GRS}{Gravitational Reference Sensor}
\newacronym{tmdws}{TM-DWS}{Test-Mass Differential Wavefront Sensing}
\newacronym{ldws}{LDWS}{Long-arm Differential Wavefront Sensing}
\newacronym[	plural={MOSAs},
		        first={Moving Optical Sub-Assembly},
		        firstplural={Moving Optical Sub-Assemblies}
            ]{mosa}{MOSA}{Moving Optical Sub-Assembly}
\newacronym{siso}{SISO}{Single-Input Single-Output}
\newacronym{mimo}{MIMO}{Multiple-Input Multiple-Output}
\newacronym[plural=MBHB's, firstplural=Massive Black Holes Binaries (MBHB's)]{mbhb}{MBHB}{Massive Black Holes Binary}
\newacronym{cmb}{CMB}{Cosmic Microwave Background}
\newacronym{sgwb}{SGWB}{Stochastic Gravitational Waves Background}
\newacronym{pta}{PTA}{Pulsar Timing Arrays}
\newacronym{gw}{GW}{Gravitational Wave}
\newacronym{snr}{SNR}{Signal-to-Noise Ratio}
\newacronym{pbh}{PBH}{Primordial Black Holes}
\newacronym{psd}{PSD}{Power Spectral Density}
\newacronym{tcb}{TCB}{Barycentric Coordinate Time}
\newacronym{bcrs}{BCRS}{Barycentric Celestial Reference System}
\newacronym{lhs}{LHS}{Left-Hand Side}
\newacronym{rhs}{RHS}{Right-Hand Side}
\newacronym{mcmc}{MCMC}{Monte-Carlo Markov Chains}
\newacronym{cs}{CS}{Cosmic Strings}
\newacronym{ssb}{SSB}{Solar System Barycentric}
\newacronym{oms}{OMS}{Optical Metrology System}
\newacronym{scird}{SciRD}{Science Requirement Document}
\newacronym{dof}{DoF}{Degree of Freedom}
